\documentclass[11pt]{article}

\usepackage{amsfonts}
\usepackage{amsmath,amssymb,amsthm,enumerate,graphicx}
\usepackage{ifthen,latexsym,syntonly}
\usepackage{setspace}
\usepackage{mathabx,bbm}
\usepackage{color}
\usepackage[round,comma,authoryear]{natbib}
\usepackage{float}
\usepackage{epstopdf}
\usepackage{mathabx}
\usepackage{relsize}
\usepackage{color}
\usepackage{calc,xspace}
\usepackage{bibentry}
\usepackage{multirow}
\usepackage{graphicx}
\usepackage{xcolor}
\usepackage{subcaption}
\usepackage[stable]{footmisc}
\usepackage{soul}
\usepackage{float}
\usepackage{booktabs}
\usepackage{caption}
\usepackage{subcaption}

\usepackage{bm}
\usepackage{algorithm,algpseudocode}

\usepackage[title]{appendix}

\usepackage{verbatim}

\bibpunct{(}{)}{,}{a}{}{,}  % for natbib
\newcommand{\mc}[1]{\mathcal{#1}}
\newcommand{\ud}{\,\mathrm{d}}
 
\DeclareMathOperator*{\argmax}{arg\,max}

\newcommand{\mb}[1]{\textcolor{blue}{\textsf{[MDB: #1]}}}

    \def\@fnsymbol#1{\ensuremath{\ifcase#1\or *\or \dagger\or \ddagger\or
    \mathsection\or \mathparagraph\or \|\or **\or \dagger\dagger
    \or \ddagger\ddagger \else\@ctrerr\fi}}

\renewcommand{\thefootnote}{\fnsymbol{footnote}}

\title{A Deep Learning Analysis of \\ Climate Change, Innovation, and Uncertainty\footnote{We are grateful for valuable suggestions and comments from Victor Duarte (discussant), Harrison Hong, Felix Kubler, Simon Scheidegger, Jose Scheinkman, and Rebecca Willett. We also thank participants at the 19th Annual Olin Finance Conference at WashU, the 2023 Conference of the Society for the Advancement of Economic Theory, the MFR Program-Bocconi University Conference on Decision-Making Under Uncertainty in Climate and Macroeconomics, and the ``Assessing the Economic and Environmental Consequences of Climate Change: Incorporating Uncertainty and Quantifying Its Importance'' Workshop at the UChicago Institute on Mathematical and Statistical Innovation (IMSI), which is supported by the National Science Foundation (Grant No. DMS-1929348). This draft is preliminary and comments are welcome.}
}

\author{Michael Barnett \\ \small{Arizona State University } \and William Brock \\ \small{University of Wisconsin} \and Lars Peter Hansen \\ \small{University of Chicago} \and Ruimeng Hu \\ \small{University of California Santa Barbara} \and Joseph Huang \\ \small{University of Pennsylvania} \and \\ \large{PRELIMINARY DRAFT} }

\date{}

\begin{document}

\maketitle

\thispagestyle{empty}

\renewcommand{\thefootnote}{\arabic{footnote}}

\begin{abstract}

%We study the implications of model uncertainty in a climate-economics setting with three types of capital: ``dirty'' capital that produces carbon emissions when used in production, ``clean'' capital that generates no emissions but is initially less productive than dirty, and knowledge capital that increases with R\&D investment and can lead to technological innovation in green sector productivity. To solve our high-dimensional, non-linear model framework we implement a neural-network-based global solution method. We show that there are first-order impacts of model uncertainty on optimal decisions and social valuations in this integrated climate-economic-innovation framework.  Specifically, accounting for interconnected uncertainty over climate dynamics, economic damages from climate change, and the arrival of a green technological change leads to significant emphasis on investment in green capital and R\&D investment in anticipation of potential technological change and the revelation of climate damage severity.

We study the implications of model uncertainty in a climate-economics framework with three types of capital: ``dirty'' capital that produces carbon emissions when used for production, ``clean'' capital that generates no emissions but is initially less productive than dirty capital, and knowledge capital that increases with R\&D investment and leads to technological innovation in green sector productivity. To solve our high-dimensional, non-linear model framework we implement a neural-network-based global solution method. We show there are first-order impacts of model uncertainty on optimal decisions and social valuations in our integrated climate-economic-innovation framework.  Accounting for interconnected uncertainty over climate dynamics, economic damages from climate change, and the arrival of a green technological change leads to substantial adjustments to investment in the different capital types in anticipation of technological change and the revelation of climate damage severity.

\end{abstract}

\thispagestyle{empty}

\newpage
\pagenumbering{arabic}
\renewcommand{\thepage} {\arabic{page}}

\section[Introduction]{Introduction}

The potential consequences of climate change are becoming increasingly apparent. The Sixth Assessment Report (AR6) of the Intergovernmental Panel on Climate Change (IPCC) states that ``[i]t is unequivocal that human influence has warmed the atmosphere, ocean and land'' and that anthropogenic climate change ``has caused widespread adverse impacts and related losses and damages to nature and people, beyond natural climate variability.''\footnote{See \cite{RN2} and \cite{RN14}.} Given these concerns, policymakers and organizations are increasingly focusing on the necessity and feasibility of a transition to a carbon-neutral economy. Reports from the OECD, the IEA, the White House, McKinsey \& Co., and Princeton University's High Meadows Environmental Institute\footnote{See \cite{oecd_report}, \cite{bouckaert2021net}, \cite{house2021long}, \cite{mckinsey2022net}, and \cite{jenkinsmission}.}, among many others, emphasize the need to heavily invest in cleaner production methods to reduce current carbon emissions, as well as the need to devote significant resources to R\&D for developing new green technology to prevent future carbon emissions. These conclusions are based largely on scenario analysis aimed at achieving climate policy goals such as the 1.5 C $^{\circ}$ GMT temperature anomaly ceiling proposed by the 2015 Paris Climate Agreement. The implementation of a socially efficient carbon-neutral transition, taking into account the various frictions, risks, uncertainties, and endogenous feedbacks and responses related to climate change and climate policy action, is the key economic question that we address in this paper. 

%Major reports from the OECD\footnote{\cite{oecd_report}}, the IEA\footnote{\cite{bouckaert2021net}}, the White House\footnote{\cite{house2021long}}, McKinsey \& Co.\footnote{\cite{mckinsey2022net}}, and Princeton University's High Meadows Environmental Institute\footnote{\cite{jenkinsmission}}, 

We develop and solve a dynamic general equilibrium framework with three types of capital: ``dirty'' capital that generates carbon emissions when used in production, ``green'' capital that produces no emissions but is initially less productive than dirty capital, and ``knowledge'' capital that increases the likelihood of a technology shock that augments the productivity of green capital. By focusing on capital stocks, our model incorporates dynamic features that play an important role in the transition to carbon-neutrality. First, the model framework leads to an emissions pathway that is ``sticky'' or persistent because dirty capital depreciates gradually and is costly to disinvest. Second, convex adjustment costs related to investment capital means that the accumulation of new green capital can take significant time. Finally, the arrival of improved green technology depends on the stock of knowledge capital in the economy, which requires R\&D investment across time and substitutes resources away from consumption, as well as from other types of investment. 

An important consideration in our analysis is the substantial uncertainty about the central mechanisms in our model. Specifically, our framework allows for uncertainty as it pertains to carbon-climate dynamics, or the mapping from carbon emissions into atmospheric carbon into temperature changes; economic damage functions, or the negative impact on output due to changes in atmospheric temperature; and technological innovation, or the likelihood of a technology shock that augments the productivity of green capital via R\&D investment. While each of these model components allows for risk in the form of stochastic realizations, we are interested in exploring alternative forms of uncertainty. This includes ambiguity about possible model parameterizations, as well as the possibility that a given model is misspecified in a meaningful and unknown way. We explicitly incorporate these uncertainty considerations into the social planner's decision problem by applying tools from dynamic decision theory. Our analysis highlights the endogenous interdependencies and feedback effects that substantially impact optimal policy choices and social valuations related to green innovation coming from uncertainty aversion.

Our general equilibrium framework with multiple endogenous state variables requires solving relatively high-dimensional PDEs with significant non-linearities due to the model uncertainty concerns and stochastic jump processes related to technological change and climate damages. As a result, our analysis requires computational methods that provide global solutions to accurately characterize the endogenous optimal policies and the social valuations of interest. Standard finite difference methods commonly used in economics and finance are not well equipped for such problems.  We, therefore, develop and implement an algorithm using deep neural network methods for our multi-dimensional, continuous-time, climate-economic framework. Our algorithm implements an extended deep Galerkin method that is able to handle multiple non-stationary, endogenous state variables with considerable non-linearity in an infinite horizon setting. Our numerical methodology is an important contribution of our work to the fields of macroeconomics, finance, and climate-economics. The algorithm provides a toolbox that considerably expands the ability of researchers to address research questions in these areas that can quickly be overwhelmed by the ``curse of dimensionality'' when accounting for regional, household, and firm heterogeneities across production technologies, economic frictions, and policy objectives.

\subsection{Climate Economics Literature}

Our paper builds on and contributes to a number of important areas of research across economics, finance, geoscience, and applied mathematics. The implications of anthropogenic emissions have been a central focus of geoscientists for many decades, beginning with the seminal work of \cite{Arrhenius:1896}. Recent work has focused on characterizing the dynamic relationship between carbon emissions, atmospheric carbon concentration, and temperature change via pulse experiments \citep{Joosetal:2013, Geoffroy:2013, Ebyetal:2009}, deriving simplified emulators or approximations of these complex relationships for the use of policymakers \citep{MatthewsGillettScott:2009, Pierrehumbert:2014, MacDougallSwartKnutti:2017}, as well as quantifying the dynamic and stochastic features of carbon-climate dynamics \citep{RickeCaldeira:2014, PalmerStevens:2019}. These components are critical inputs into our model framework and uncertainty analysis.

The climate economics literature has given substantial focus to estimating the economic consequences of climate change and deriving a Social Cost of Carbon (SCC), or the cost to social welfare of emitting an additional amount of carbon emissions into the atmosphere. Economists have estimated the economic costs generated by observed climate change for numerous economic sectors, regions, and dimensions of the economy \citep{dell2012temperature, Hsiangetal:2017, colacito2019temperature, IPCC:2019, CareltonGreenstone:2021}. Theoretical modeling and integrated assessment model analysis to derive SCC valuations have considered for numerous economic mechanisms \citep{Golosovetal:2014, Acemogluetal:2016, Nordhaus:2018, CaiJuddLontzek:2017, CaiLontzek:2019} and proposed climate damage function approximations and frameworks \citep{Lentonetal:2008, Weitzman:2012, Caietal:2015, Drijfhoutetal:2015, Ritchieetal:2021}. 

Recent work has begun to examine important dimensions related to interconnected climate change and economic model uncertainty \citep{Olsonetal:2012, LemoineTraeger:2014, HasslerKrusellOlovsson:2018, Nordhaus:2018, DietzVenmans:2019, BarnettBrockHansen:2020, Rudik:2020, BarnettBrockHansen:2021, barnett2023climate}. Importantly, many of these studies of uncertainty have exploited the powerful toolset developed in dynamic decision theory \citep{AndersonHansenSargent:2003, MaccheroniMarinacciRustichini:2006, HansenSargent:2007, KlibanoffMarinacciMukerji:2009, HansenMiao:2018, BarnettBrockHansen:2020, Cerreia-Vioglioetal:2021}, allowing researchers to account for model uncertainty explicitly in the decision-maker's problem within the model.

This paper also links to an important area of macroeconomic research. This includes the foundational work on endogenous economic growth \citep{brock1972optimal, baumol1986productivity, lucas1988mechanics, romer1990endogenous}, as well as recent analysis of the social and private benefits of innovation \citep{bloom2019toolkit, lucking2019have}, and considerations for the transition to a green economy \citep{Acemogluetal:2012, Acemogluetal:2016, jaakkola2019non}. In addition, the connection between macroeconomics and asset pricing in production-based asset pricing \citep{brock1982asset, cochrane1991production, jermann1998asset}, as well as the importance links to economic growth and innovation \citep{papanikolaou2011investment, kogan2014growth, kung2015innovation} have important implications for the social valuations we derive in our climate-economics-innovation linked framework.

\subsection{Deep Learning Literature}

In recent years, deep learning algorithms, built on the neural network's remarkable ability to represent and approximate high-dimensional functions and efficient gradient descent optimizers, have been very successful in many areas, ranging from computer vision and speech recognition to scientific computing (see, e.g. \cite{CaTr:17,GaDu:17,HaJeE:18,ZhHaE:18,han2019uniformly}). The climate change problem considered in this paper aims to study how choices for investment in clean and dirty capital, and R\&D for technological innovation, are determined when facing uncertainties, which, mathematically, is a stochastic control problem. Along this direction, deep learning algorithms are roughly categorized into: direct parameterization, partial differential equations (PDEs), and forward-backward stochastic differential equations (FBSDEs) approach.

In this first category, the seminal work in high-dimension was proposed by \cite{han2016deep}, which solves a global-in-time minimization problem in accordance with the utility of the control problem. Later, \cite{bachouch2021deepnumerical} extended to the local-in-time approach combined with dynamic programming techniques. \cite{han2020rnn} solve the control problem with aftereffects, modeled by stochastic differential delayed equations, using the same spirit of \cite{han2016deep} but with advanced network architectures, and \cite{carmona2022convergence} extended to mean-field control problems. 

In the PDE approach, the stochastic control problem will first be reformulated into a Hamilton-Jacobi-Bellman (HJB) PDE using the dynamic programming principle, and then solved by deep learning algorithms. For generic PDEs, \cite{SiSp:18} proposed the deep Galerkin method (DGM), and \cite{raissi2019physics} proposed the physics-informed neural networks (PINNs), roughly at the same time. Both approximate solutions to PDEs by training neural networks to minimize the residuals coming from the initial conditions, the boundary conditions, and the PDE operators. \cite{saporito2021path} followed this idea and solved path-dependent PDEs using recurrent neural networks. Later, \cite{al2022extensions} extended the DGM to deal with HJB equations in their unsimplified primal form and solved for the value function and the optimal control simultaneously by characterizing both as deep neural networks.

For semi-linear parabolic PDEs, which correspond to stochastic control problems with uncontrolled volatility, \cite{MR3736669} and \cite{HaJeE:18} proposed a deep BSDE solver which, to the best of our knowledge, is the first work in this area and has inspired much follow-up work. They recast the solution of a semi-linear parabolic PDE into a global-in-time optimization problem based on the associated BSDEs via the non-linear Feynman Kac formula and variational form. \cite{hurephamwarin2020deep} dealt with the same associated BSDE but obtained the resolution by solving backward in time using a sequence of small optimization problems and termed it as deep learning backward dynamic programming (DBDP). Another work focusing on semi-linear PDE is the deep splitting method proposed by \cite{beck2021deep}, where the partial differential operators are split into the linear part and the nonlinear part, with the nonlinear part propagating first by freezing the solution and the linear part then being taken care of by an approximate Feynman-Kac representation. When the control appears in the state dynamics' diffusion coefficient, this leads to a fully nonlinear PDE. In this direction, in the same spirit of \cite{MR3736669} and \cite{HaJeE:18}, \cite{BeEJe:19} proposed a DL algorithm by solving the corresponding second-order BSDE, and \cite{pham2021neural} extended the DBDP idea in \cite{hurephamwarin2020deep} by further approximating the Hessian matrix using auto differentiation of first derivatives' neural networks.

In the FBSDE approach, the control problem will first be reformulated into a coupled forward-backward system using the stochastic Pontryagin maximum principle. The system is, in general, hard to solve numerically, let alone in high dimensions, due to its coupled nature and opposing directions: one with an initial condition running forward in time and one with a terminal condition running backward in time. \cite{han2020convergence} solved the fully-coupled FBSDE by extending the deep BSDE solver with convergence analysis subject to neural networks' universal approximation, and \cite{ji2020three} further extended the results with the Picard iteration method.

Recently, attention has also been drawn to stochastic control problems with jumps. The problem is more involved as the PDE becomes partial integro-differential, thus non-local; and the FBSDE system now contains a L\'{e}vy process. 
For such problems, \cite{boussange2022deep} solved the associated non-local PDE by extending the deep splitting method in \cite{beck2021deep} and multilevel Picard approximation  method in \cite{hutzenthaler2019multilevel}; \cite{castro2022deep} solved the associated forward-backward system by extending the DBDP method in \cite{hurephamwarin2020deep}; and \cite{gnoatto2022deep} solved the same system in the same spirit of the deep BSDE solver, proposed in \cite{MR3736669} 
 and \cite{HaJeE:18}.

When using neural networks to parameterize the quantity of interest, usually the value function or the control policy, people may want to incorporate domain knowledge, such as guaranteed monotonicity or convexity of the learned function with respect to some of the input variables. There have been fruitful studies on how to build network structures in order to fulfill such requirements, see for instance, \cite{sill1997monotonic}, \cite{gupta2016monotonic}, \cite{wehenkel2019unconstrained} and \cite{runje2022constrained}.

%%%%%%%%%%%%%%%%%%%%%%%%%%%%%%%%%%%%%%%%%%%%%
%%%%%%%%%%%%%%%%%%%%%%%%%%%%%%%%%%%%%%%%%%%%%

\section{Climate-Economics Model}

We now outline the main model for our analysis. The model incorporates components related to climate dynamics, economic elements and dynamics related to damages from climate change, production technologies, preferences, and technological innovation through R\&D, as well as model uncertainty aversion. We outline each of these model pieces in what follows, and then derive a number of theoretical results before presenting the numerical solutions to the model. 

\subsection{Climate Dynamics}

Recent work in geosciences has focused on constructing simplified approximations of climate dynamics generated by large-scale atmosphere-ocean general circulation models (AOGCM) for use in policy analysis. These include a tractable framework demonstrated by \cite{MatthewsGillettScott:2009} and \cite{Friedlingstein:2015} as a proportional relationship between temperature change and cumulative carbon emissions. This relationship is of the general form:
\begin{eqnarray*}
\textrm{ temperature anomaly }  \textrm{ } \approx \textrm{\underline{climate sensitivity}} \textrm{ } (\beta_f) \textrm{ }  \times \textrm{ cumulative emissions.}
\end{eqnarray*}

\begin{figure}[!pht]
\centering
\includegraphics[width=.85\textwidth]{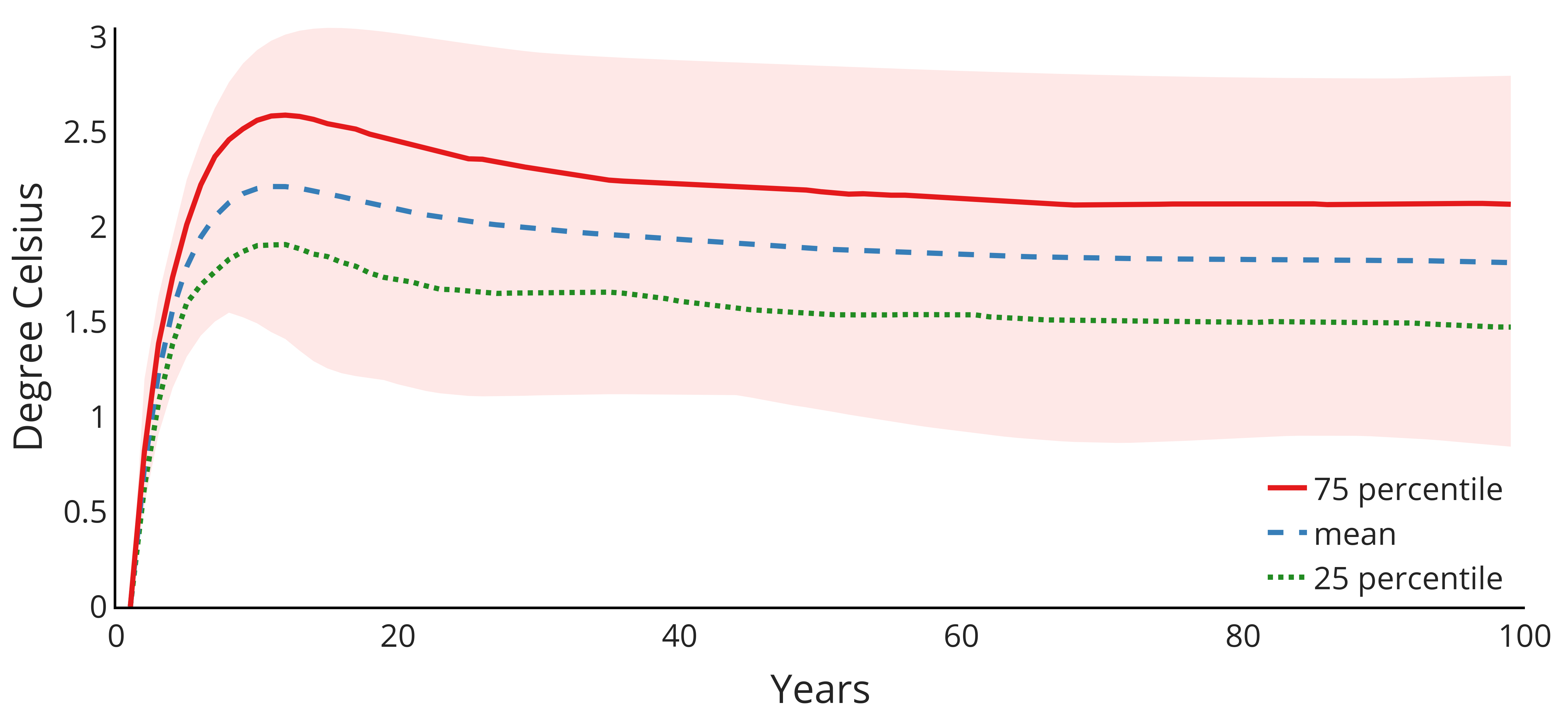} \\
\includegraphics[width=.7\textwidth]{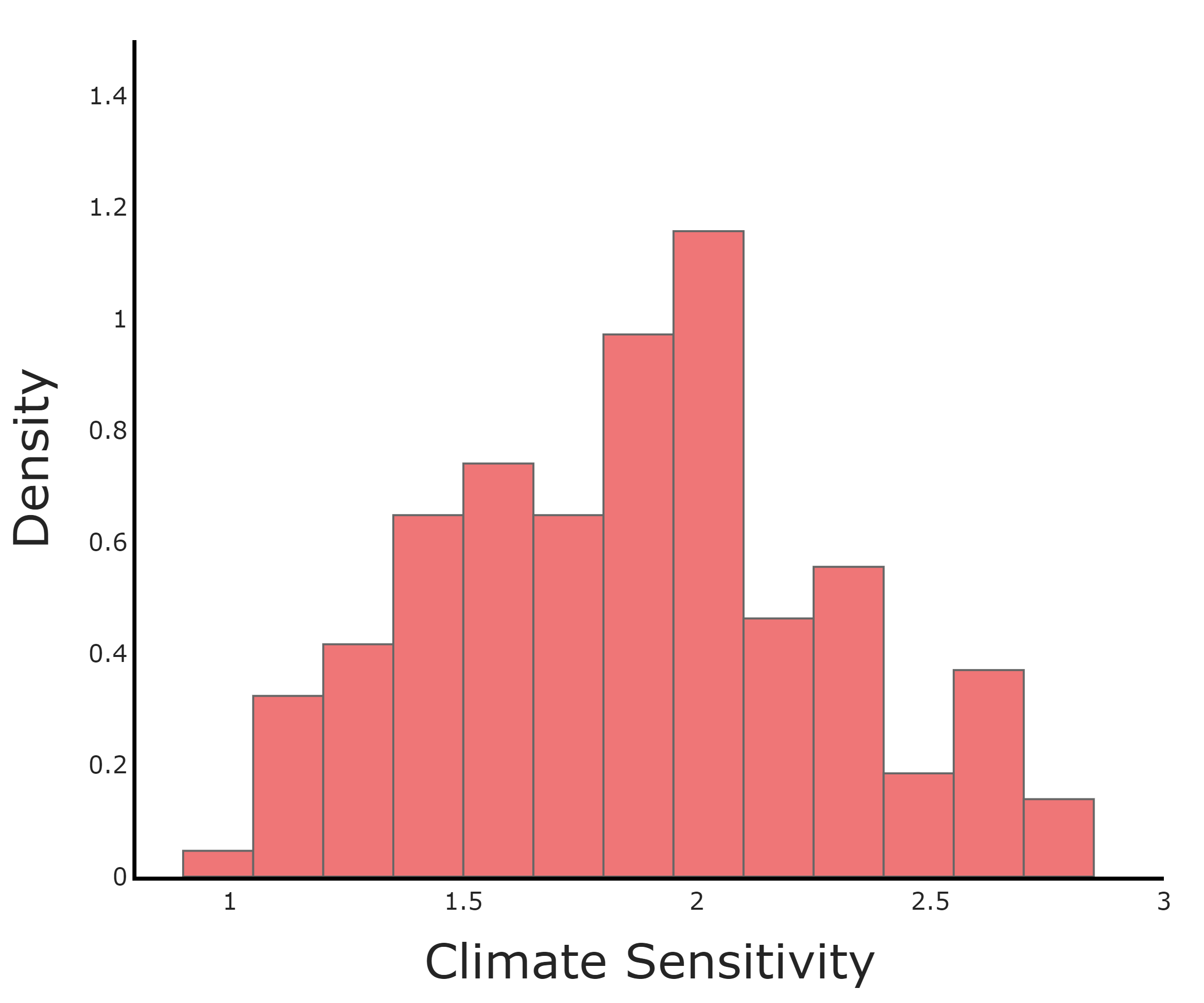}
\caption{Implied TCRE coefficients across time and models, capturing carbon and temperature uncertainty. The top figure shows the percentiles for temperature responses to emission pulses for all carbon and temperature model combinations. The bottom figure is a histogram for the exponentially weighted average responses of temperature to an emissions pulse from 144 different models  using  a rate $\delta = .01$.}\label{fig:TCRE}
%\parbox{\textwidth}{\footnotesize Percentiles for temperature responses to emission pulses for all carbon and temperature model combinations. Captures carbon and temperature uncertainty.}
\end{figure}

Not only is this approximate relationship straightforward to incorporate into most continuous-time macroeconomic and asset pricing models, it also directly provides a model comparison set to use in our uncertainty analysis based on the constant of proportionality implied by different large-scale climate models. For our analysis, we use a stochastic variant of this relationship 
\begin{eqnarray*}
dY_t = E_t(\beta_{f}  dt + \varsigma dW_t),
\end{eqnarray*}
where $Y_t$ is the global mean temperature anomaly, with respect to the preindustrial level, $E_t$ is global carbon emissions, $W_t$ is a Brownian motion process with filtration $\mathcal{F}_t$, $\varsigma$ is the volatility loading for temperature, and $\beta_f$ is the climate sensitivity proportionality constant. This model abstracts from transitory ``weather'' fluctuations in temperature and assumes that emissions today have a long-lasting, i.e., permanent, impact on future temperature. The inclusion of stochastic variation, which allows for a meaningful uncertainty analysis by ``disguising'' the climate model in the statistical sense that a small set of observations do not immediately reveal the true model, is motivated by the argument about climate model predictability put forward in \cite{PalmerStevens:2019}. The parameter $\beta_f$ varies across different carbon-climate dynamics models, and we will denote these different values as $\beta_{f,\ell}$ for a given model $\ell \in \{1, ..., \mathcal{L} \}$. In the geoscience literature, this parameter is known as the transient climate response to cumulative emissions (TCRE) parameter and serves as a climate sensitivity measure.

While the proportionality relationship used here is based on a relationship of carbon emissions to temperature change that is born out at a time horizon of 10 years and beyond, \citep{DietzVenmans:2019, BarnettBrockHansen:2021, barnett2023climate} and others have shown that this relationship is well suited for frequencies as short as one year. We follow \cite{BarnettBrockHansen:2021} and use pulse experiment results of \cite{Joosetal:2013} and \cite{Geoffroy:2013} to build the set of climate sensitivity models for our analysis. \cite{Joosetal:2013} examines carbon dynamics variation and uncertainty based on responses of atmospheric carbon concentration to emission pulses of 100 GtC for several alternative Earth System models. \cite{Geoffroy:2013} provides temperature dynamics variation and uncertainty based on approximate dynamics relating the log of atmospheric carbon to future temperature, following \citeauthor{Arrhenius:1896} equation. We combine these two sets of pulse experiments by taking 9 different atmospheric carbon responses as inputs into 16 temperature dynamics approximations to derive 144 different carbon-climate TCRE models. Figure \ref{fig:TCRE} shows the dynamic pathways and the implied TCRE parameters for these 144 models. We can see from the pathways that the temperature response peaks around 10 years, and flattens out thereafter, and the implied TCRE parameters are quite dispersed, with values ranging from around 1 to just below 3, with the average value being $1.86$.

\subsection{Climate Damages}\label{section:damage_function}

\begin{figure}[!pht]
%\caption{Climate Change Model Uncertainty} \label{fig:climate_uncertainty}
\begin{center}
            \includegraphics[width=0.95\textwidth]{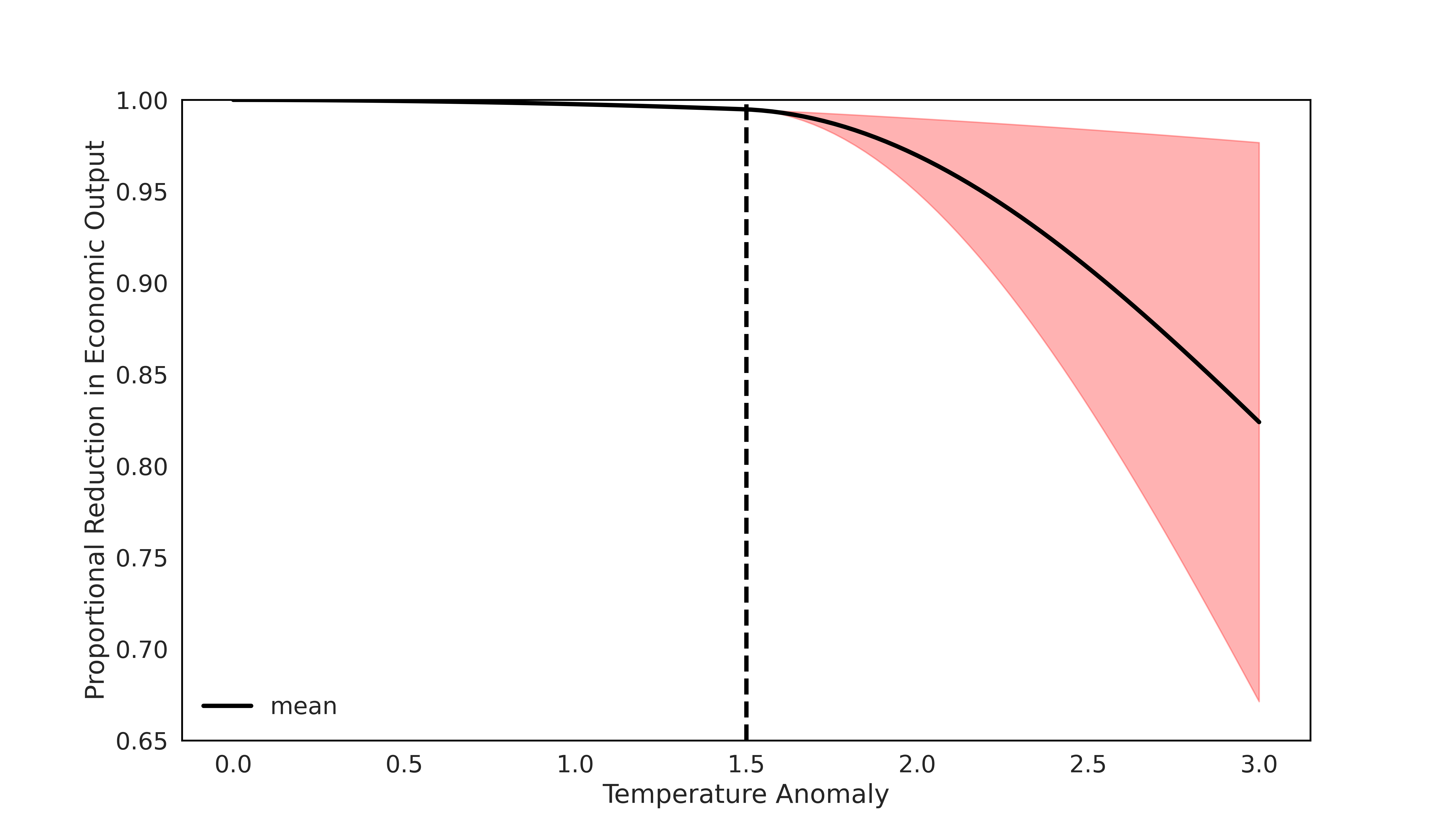}  \\             \includegraphics[width=0.95\textwidth]{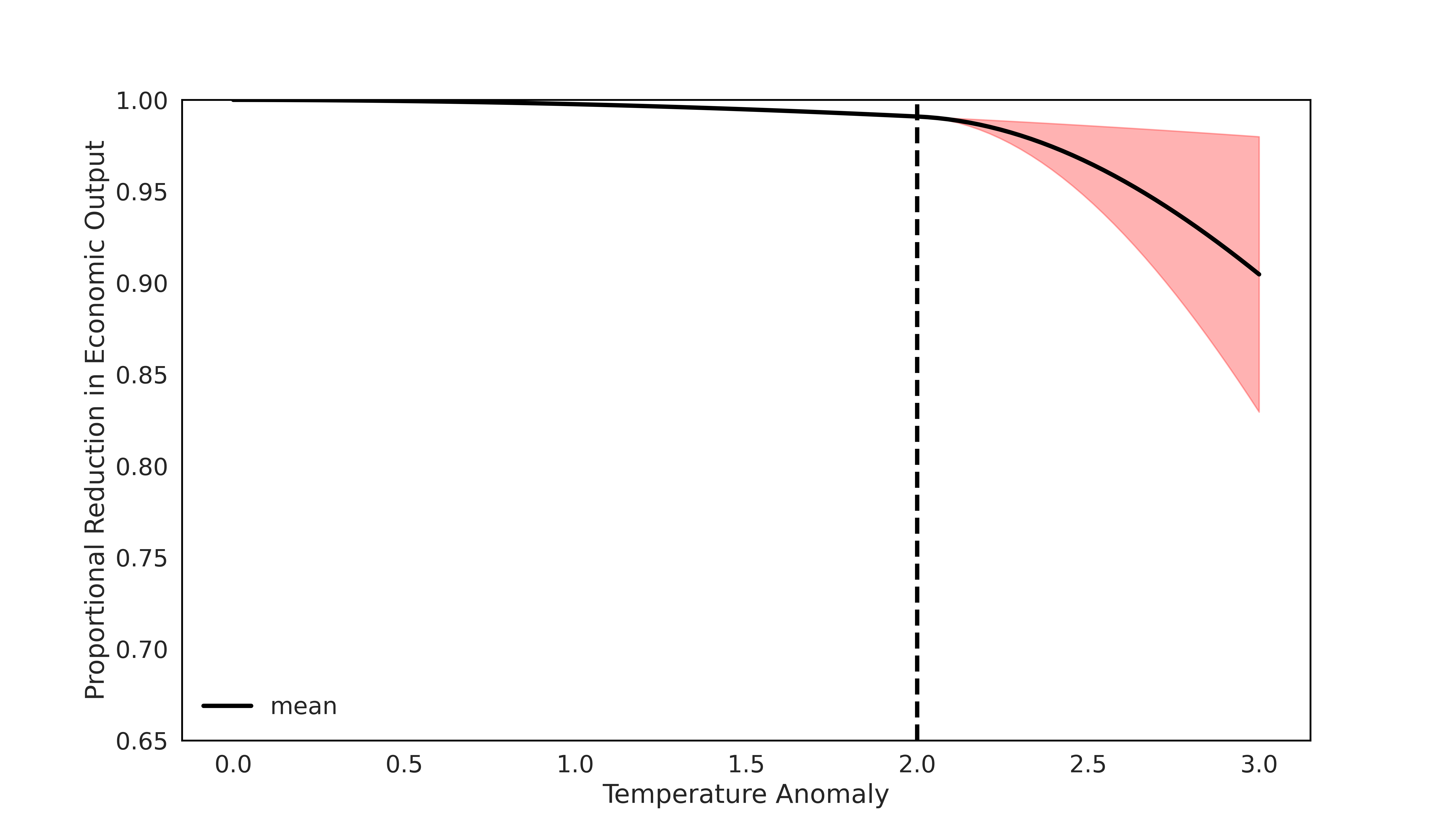}
\end{center}        
\caption{Range of possible damage functions for different jump thresholds for two cases. The shaded region in each plot gives the range of possible values for exp(-n), which measures the proportional reduction of the productive capacity of the economy. The top figure shows the damage function curvature when the jump occurs at $Y_t = {\bar y} = 1.5$. The bottom figure shows the damage function curvature when the jump occurs at $Y_t = {\underline y} = 2.0$.}\label{fig:climate_damages}

\end{figure}

Following \cite{BarnettBrockHansenZhang:2023}, our model of economic damages from climate change, or climate damages, is piece-wise log quadratic. Damages $N_t$ are assumed to impact capital, output, investments, and consumption in a proportional manner. This specification allows for a change in the evolution of damages that is triggered by a jump process realization. Uncertainty about climate damages is connected to the values of ${\hat Y}$, which is defined by ${\hat Y} = Y + {\bar y} - {\tilde Y}$ for ${\underline y} < {\tilde y} \le {\bar y}$, and $\gamma_3^m$, for which there are alternatives values denoted by $m \in \mathcal{ M}$. There is a stochastic jump process that determines dynamically which value of $\gamma_3^m$ will be realized for the damage function and at the value of $\tilde{Y}$ the jump will be realized. Specifically, the jump process has $m$ absorbing states, where each state $m$ corresponds to a value of $\gamma_3^m$. Before the realization of the jump, each $\gamma_3^m$ value has a prior probability $\pi_m^p$. When the jump is realized, the value of $\gamma_3^m$ is revealed and the true damage function curvature becomes known to the social planner in the model. The likelihood of a damage function jump is determined by a jump intensity $\mathcal{I}_d(y)$ that is increasing in the endogenous temperature anomaly state variable $y$ and is localized to the range of temperature values $[\underline{y},\overline{y}]$. At the time of the jump $\tau$, the state variable ${\hat Y}$ shifts from the temperature anomaly $Y = \tilde{Y}_\tau$ to the damage jump threshold ${\bar y}$. Thus $\hat{Y}$ is a state variable that shifts from capturing just the temperature anomaly to a jump-adjusted measure capturing the derivative shift of the damage function as well. The uncertainty is associated both with the probability distribution on possible $\gamma_3^m$ model realizations, as well as the probability of when a damage jump will occur in the $[\underline{y},\overline{y}]$. The full functional form is left for the Appendix, but the evolution of log damages is given by
\begin{equation*} \label{damage_function}
d \log  N_t  = \left\{ \begin{array}{ll} \left(\gamma_1 + \gamma_2 {\widehat Y}_t \right)  E_t \left( \beta_{f,\ell} dt + \varsigma dW_t\right)  +  \frac {\gamma_2 |\varsigma|^2 \left(E_t\right)^2}{2} dt,  & t  \le \tau, \\
\left(\gamma_1 + \gamma_2 {\widehat Y}_t + \gamma_3^m \left({\widehat Y}_t - {\bar y}\right) \right)  E_t \left( \beta_{f,\ell} dt + \varsigma dW_t\right)  +  \frac {\left( \gamma_2 + \gamma_3^m \right) |\varsigma|^2 \left(E_t\right)^2}{2} dt,   & t  \le \tau. \end{array} \right.  
\end{equation*}

For our main setting, we assume $[\underline{y},\overline{y}] = [1.5, 2.0]$, consistent with the critical temperature thresholds used by the IPCC and the 2015 Paris Climate Agreement. Figure \ref{fig:climate_damages} shows the implied climate damages across all $\gamma_3^m$ values for the jump realization occurring at the temperature anomaly values of ${\tilde y} = \underline{y} = 1.5$ and ${\tilde y} = \bar{y} = 2$ which bound our range for the damage jumps occurring. The red shaded region shows the range of possible damage outcomes across the possible values of $\gamma_3^m$, and the black line gives the mean value.\footnote{While these plots are nearly identical to those found in \cite{BarnettBrockHansen:2021}, there are key differences. Specifically, in \cite{BarnettBrockHansen:2021}, there is a possibility for discontinuities in the damage function and jumps in the climate damages, which are ignored in their analysis. Our specification removes such discontinuities, and allows only for smooth shifts in the derivative of the damage function.}  While there is significant discussion about the likelihood of tipping points or thresholds at the global scale captured by discrete jumps or shifts in climate damages\footnote{See \cite{Brooketal:2013} and  \cite{Levitan:2013} for examples on this discussion.}, our setting is instead a smooth shift where the planner realizes whether the damage function is more concave than previously known. In this sense, we view our framework as a valuable tool for characterizing the possibility and uncertainty of potentially severe, nonlinear climate damage outcomes that are highly relevant for the optimal policy decisions of a social planner confronting climate change. Moreover, the structure of our uncertainty is novel with regards to much of the existing climate-economics literature, as the information dynamics in most settings is static, or uncertainty about the damage function is never resolved.

%%%%%%%%%%%%%%%%%%%%%%%%%%%%%%%%%%%%%%%%%%%%%%%%

\subsection{Production}

For the economic component of the model, we assume there are two sectors producing perfectly substitutable consumption goods using their own AK technologies: 
\begin{eqnarray*}
F_i(K_{i,t}) & = & A_{i} K_{i,t}, i \in \{d,g\} , \\
F(K_{d,t},K_{g,t}) & = & F_d(K_{d,t}) + F_g(K_{g,t}) = A_{d} K_{d,t} + A_{g} K_{g,t} .
\end{eqnarray*}

Each sector has its own capital stock that evolves with logarithmic adjustment costs and Brownian shocks as follows:
\begin{align*}
dK_{d}/K_{d} & =[\alpha_{d}+\Gamma_d \log (1+ \theta_d i_{d})]dt+\sigma_{d}dW , \\
dK_{g}/K_{g} & =[\alpha_{g}+\Gamma_g \log (1+ \theta_g i_{g})]dt+\sigma_{g}dW .
\end{align*}

For computational tractability, we redefine the state variables characterizing the capital stocks in the model and use $\log K=\log(K_{d}+K_{g})$ and $R=\frac{K_{g}}{K_{d}+K_{g}}$. By Ito's lemma, these two state variables have the following evolution processes:
\begin{align*}
d\log K & =(1-R)[\alpha_{d}+\Gamma_d \log(1+ \theta_d i_{d})]dt+R[\alpha_{g}+\Gamma_g \log(1+ \theta_g i_{g})]dt\\
 & -\frac{1}{2}|\sigma_{d}(1-R)|^{2}-\frac{1}{2}|\sigma_{g}R|^{2}dt\\
 & +(1-R)\sigma_{d}dW_{d}+R\sigma_{g}dW_{g}, \\
dR/R & =(1-R)\{[\alpha_{g}+\Gamma_g \log(1+ \theta_g i_{g})]-[\alpha_{d}+\Gamma_d \log(1+ \theta_d i_{d})]\}dt\\
 & +(1-R)\{(1-R)|\sigma_{d}|^{2}-R|\sigma_{g}|^{2}\}dt\\
 & +(1-R)\sigma_{g}dW_{g}-(1-R)\sigma_{d}dW_{d}.
\end{align*}

There are two key differences across these different consumption good production sectors in the economy. First, Sector d is the only sector that generates emissions, so that $E_t$ in our temperature evolution equation is given by 
\begin{eqnarray*}
    E_{t}=\eta A_{d}K_{d} ,
\end{eqnarray*}
where $\eta$ is the emissions intensity parameter of Sector d production. Second, Sector d is initially more productive than Sector g in that $A_d > A_g$. While Sector d is initially more productive than Sector g, we also assume that there is the potential for a ``green'' technology shock that would augment Sector g productivity. The arrival rate of this one-time jump in Sector g productivity is an increasing function of the aggregate knowledge capital stock in the economy $\kappa$. The evolution of the aggregate knowledge stock is given by
\begin{align*}
d\log\kappa & =-\zeta dt+\psi_{0}(\frac{I_{\kappa}}{\kappa})^{\psi_{1}} dt-\frac{|\sigma_{\kappa}|^{2}}{2}dt+\sigma_{\kappa}\cdot dW_{t}, 
\end{align*}
where $I_{\kappa}$ is the level of R\&D investment made to increase the knowledge capital stock, $\zeta$ is the decay rate of the knowledge stock, $\psi_0$ and $\psi_1$ capture the effectiveness of R\&D investment, and $\sigma_{\kappa}$ is the volatility associated with the knowledge capital stock. 

We include uncertainty about the realization of the technology shock in the form of a discrete set of possible realizations for the post-technology jump Sector g productivity $A_g^j$. The dynamic realization of the technological change shock that determines the value of $A_g^j$ is through a stochastic jump process. There are $j$ absorbing states, where alternative values are denoted by $j \in \mathcal{J}$, with each state $j$ corresponding to a particular value of $A_g^j$ and each value of $A_g^j$ having a prior probability $\pi_g^j$. For each potential realization of the technology shock, the value of $A_g^j$ is such that $A_g^j \ge A_g$. The likelihood of a technological change jump is dependent upon a jump intensity $\mathcal{I}_g(\kappa)$. We choose the simple functional form $\mathcal{I}_g(\kappa) = \kappa \varrho$, where $\kappa > 0$ is a constant, so that the arrival rate is increasing in the endogenous knowledge stock state variable $\kappa$. The information dynamics and framework and structure of the technological change jump are similar to the damage function jump. Once the jump is realized, the true technological change outcome for the value of $A_g^j$ is known to the social planner. In addition, the uncertainty pertains to the probability distribution of potential $A_g^j$ realizations and the arrival rate of the technological change shock. This structure focuses on the economic implications related to the uncertainty of breakthrough green technologies, again a relatively novel channel of innovation compared to the climate-economics literature. We find this specification appealing as it allows for a broader interpretation of the uncertainty of ``green'' technology change that further enriches our analysis and uncertainty quantification as it relates to optimal policy considerations related to technological innovation and climate change.

\subsection{Preferences}

Finally, we assume that flow utility in the model is a log function over damaged aggregate consumption, where exponential-quadratic damages multiplicatively scale consumption. The utility function is therefore given by
\begin{align*}
U(\tilde{C}) = U(C/N) = \delta\log(A_{d}K_{d}-i_{d}K_{d}+A_{g}K_{g}-i_{g}K_{g}-i_{\kappa} (K_g + K_d))-\delta\log N,
\end{align*}
where $\delta$ is the subjective discount rate, the two types of consumption goods are perfectly substitutable. Investment decisions are made by optimally dividing the output net of consumption across investment into the three types of capital in the economy (dirty production capital, green production capital, and knowledge capital), so that the market clearing final output constraint is given by the relationship
\begin{align*}
C & =A_{d}K_{d}-i_{d}K_{d}+A_{g}K_{g}-i_{g}K_{g}-i_{\kappa} (K_g + K_d),
\end{align*}
where $i_{d} = I_{d}/K_{d}$ is the dirty investment-to-dirty capital ratio, $i_{g} = I_{g}/K_{g}$ is the green investment-to-green capital ratio, and $i_{\kappa} = I_{\kappa}/(K_d + K_g)$ is the R\&D investment-to-total capital ratio.

\section{Model Solution}

With the model framework laid out, we can not turn to solving the Hamilton-Jacobi-Bellman equations that characterize the solution to the Social Planner's problem in our model. The solution to our model is a recursive Markovian equilibrium that solves the Hamilton-Jacobi-Bellman equation characterizing the planner's social welfare function, or value function. Therefore, the equilibrium solution is determined by optimal investment choices: 
\begin{eqnarray*}
\{ i_{g}^*, i_{d}^*, i_{\kappa}^* \}
\end{eqnarray*} 
that maximize the planner's discounted, lifetime expected utility as functions of the state variables 
\begin{eqnarray*}
\{\log K,R,Y_{t},\log\kappa,\log N_{t}\}.    
\end{eqnarray*} 
These optimal controls must satisfy the evolution equations for the state variables, as well as the market clearing conditions given in the model set-up above and the first order conditions from the HJB equations we outline below. To arrive at the full model solution, we must derive solutions sequentially working from the post-technology jump, post-damage-function-jump state back to the pre-technology jump, pre-damage-function-jump state, accounting for the different possible combinations of intermediate states, which are the pre-technology jump, post-damage-function-jump state and the post-technology jump, pre-damage-function-jump state. %We outline the HJB equation for each state in what follows.

\subsection{Post Damage and Technology Jumps HJB Equation}

The first jump-state we must solve is after the technology jump and damage jump have already occurred, or the post-technology jump, post-damage-function-jump state. We denote this value function by $V^{(m,j)}(K_{d},K_{g},\hat{Y},\log N)$, indicating that we are at a given realization of $\gamma_3^m$ and in the second technology state with green capital productivity $A_g^j$. We can guess-and-verify that the value function solution in this stage can be analytically simplified as follows
\begin{align*}
V^{(m,j)}(K_{d},K_{g},\hat{Y},\log N)=v^{(m,j)}(\log K,R,\hat{Y})-\log N_{t}.
\end{align*}
The remaining HJB equation characterizing the simplified value function is given by:
\begin{equation*}
\begin{split}
\delta v^{(m,j)} & =\max_{i_{g},i_{d}} \delta\log([A_{d}-i_{d}](1-R)+[A_{g}'-i_{g}]R -i_{\kappa}) + \delta\log K\\
 & + \left((1-R) \left[\alpha_{d}+\Gamma_d \log (1+ \theta_d i_{d}) \right]+R \left[\alpha_{g}+\Gamma_g \log (1+ \theta_g i_{g})\right]-\frac{   \sigma_d^2 (1-R)^2 + \sigma_g^2 R^2 }{2}  \right)v^{(m,j)}_{\log K}\\
 &+ \frac{  \sigma_d^2 (1-R)^2 + \sigma_g^2 R^2}{2} v^{(m,j)}_{\log K,  \log K}  \\
 & +\left( \left[\alpha_{g}+\Gamma_g \log (1+ \theta_g i_{g})\right] - R \sigma_g^2 - \left[\alpha_{d}+\Gamma_d \log (1+ \theta_d i_{d})\right] + (1-R) \sigma_d^2
  \right) R (1-R) v^{(m,j)}_{R}\\
 & + \frac{1}{2} R^2 (1-R)^2 ({ \sigma_g^2 + \sigma_d^2} ) v^{(m,j)}_{RR} \\
  & + \left[-R(1-R)^2\sigma_d^2 + R^2(1-R)\sigma_g^2\right]v^{(m,j)}_{\log K, R} \\
 & +\beta_{f}\eta A_{d}(1-R)K v^{(m,j)}_{y}+\frac{|\varsigma|^{2}(\eta A_{d}(1-R)K)^{2}}{2} v^{(m,j)}_{yy}\\
 & - \left( \{\gamma_{1}+\gamma_{2}\hat{Y}  + \gamma_3^m (\hat{Y} - \underbar{y}) \}\beta_{f}\eta A_{d}(1-R)K+\frac{1}{2}(\gamma_{2} + \gamma_3^m ) |\varsigma|^{2} \eta A_{d}(1-R)K)^{2} \right).
\end{split}
\end{equation*}

Taking first order conditions with respect to $i_d$ and $i_g$, we end up with the following two expressions for optimal investment choices
\begin{align*}
\delta([A_{d}-i_{d}](1-R)+[A_{g}'-i_{g}]R)^{-1}	& = \Gamma_{d}\theta_{d}(1+\theta_{d}i_{d})^{-1}[v^{(m,j)}_{\log K}-Rv^{(m,j)}_{R}], \\
\delta([A_{d}-i_{d}](1-R)+[A_{g}'-i_{g}]R)^{-1}	 & = \Gamma_{g}\theta_{g}(1+\theta_{g}i_{g})^{-1}[v^{(m,j)}_{\log K}+(1-R)v^{(m,j)}_{R}]. 
\end{align*}
Each expression equates the marginal benefit of an additional unit of a given type of capital to the marginal utility of consumption, which is the utility loss of forgoing consumption for an additional unit of investment. 

\subsection{Intermediate Jump State HJB Equations}

Our next two HJB equations come from the intermediate jump states. The first case is when the technology jump has already occurred, but the damage function jump has not. In this case, we can apply a similar analytical simplification to the value function for the social planner. We denote this value function by $V^{(j)}(K_{d}, K_{g,}Y, \log N)$, indicating that we have not had a realization of $\gamma_3^m$, but we are in the second technology state with green capital productivity $A_g^j$. The value function in this stage is given by
\begin{align*}
V^{(j)}(K_{d},K_{g,}Y,\log N)=v^{(j)}(\log K,R,Y_{t})-\log N_{t}.
\end{align*}

There are two adjustments made to the simplified HJB equation in this case relative to the post-technology and -damage jump case. First, because the damage jump has not yet occurred, the equation capturing the evolution of damages is altered to be
\begin{equation*}
\begin{split}
  & - \left( \{\gamma_{1}+\gamma_{2}Y )\}\beta_{f}\eta A_{d}(1-R)K+\frac{1}{2}(\gamma_{2} ) |\varsigma|^{2}(\eta A_{d}(1-R)K)^{2} \right). 
\end{split}
\end{equation*}
Second, the expectation of damage function jump introduces the additional term
\begin{equation*}
\begin{split}
\mathcal{I}_d(y)  \sum_{m=1}^{M} \pi_d^m (v^{(m,j)} - v^{(j)}).
\end{split}
\end{equation*}

We again take first order conditions with respect to $i_d$ and $i_g$. These expressions are unchanged from the post-damage and technology jump case, equating the marginal benefit of an additional unit of a given type of capital to the utility loss of forgoing consumption for an additional unit of investment, which is the marginal utility of consumption. 

%\subsubsection{Pre-Damage-Jump, Post-Technology-Jump Model solution}

\begin{comment}
The value function solution in this stage is given by
\begin{align*}
V(K_{d},K_{g,}Y,\log N)=v(\log K,R,Y_{t})-\log N_{t}
\end{align*}
where the remaining HJB equation characterizing the simplified value function is given by:
\begin{equation}
\begin{split}
\delta V & =\max_{i_{g},i_{d}}\delta\log([A_{d}-i_{d}](1-R)+[A_{g}'-i_{g}]R -i_{\kappa}) + \delta\log K\\
 & + \left((1-R) \left[\alpha_{d}+\Gamma_d \log (1+ \theta_d i_{d}) \right]+R \left[\alpha_{g}+\Gamma_g \log (1+ \theta_g i_{g})\right]-\frac{   \sigma_d^2 (1-R)^2 + \sigma_g^2 R^2 }{2}  \right)v_{\log K}\\
 &+ \frac{  \sigma_d^2 (1-R)^2 + \sigma_g^2 R^2}{2} v_{\log K,  \log K}  \\
 & +\left( \left[\alpha_{g}+\Gamma_g \log (1+ \theta_g i_{g})\right] - R \sigma_g^2 - \left[\alpha_{d}+\Gamma_d \log (1+ \theta_d i_{d})\right] + (1-R) \sigma_d^2
  \right) R (1-R) v_{R}\\
 & + \frac{1}{2} R^2 (1-R)^2 ({ \sigma_g^2 + \sigma_d^2} ) v_{RR} \\
  & + \left[-R(1-R)^2\sigma_d^2 + R^2(1-R)\sigma_g^2\right]v_{\log K, R} \\
 & +\beta_{f}\eta A_{d}(1-R)K v_{y}+\frac{|\varsigma|^{2}(\eta A_{d}(1-R)K)^{2}}{2} v_{yy}\\
 & - \left( \{\gamma_{1}+\gamma_{2}Y \}\beta_{f}\eta A_{d}(1-R)K+\frac{1}{2}\gamma_{2} |\varsigma|^{2}(A_{d}(1-R)K)^{2} \right) \\
 & + \mathcal{I}_d(y)  \sum_{m=1}^{M} \pi_d^m (v^{(m)} - v)
\end{split}
\end{equation}
\end{comment}

%\subsubsection{Post-Damage-Jump, Pre-Technology-Jump Model solution}

The second intermediate case is when the technology jump has not yet occurred, but the damage function jump has. We can again apply the same analytical simplification to the value function for the social planner. We denote this value function by $V^{(m)}(K_{d}, K_{g,}Y,\log \kappa, \log N)$, indicating that we have had a realization of $\gamma_3^m$, but we are in the technology state with green capital productivity $A_g$. The value function in this stage is given by
\begin{align*}
V^{(m)}(K_{d},K_{g},\hat{Y},\log \kappa,\log N)=v^{(m)}(\log K,R,\hat{Y},\log \kappa)-\log N_{t}.
\end{align*}

Relative to the post-technology and -damage jump case, we introduce the following new terms to the simplified HJB equation 
\begin{equation*}
\begin{split}
& (-\zeta+\psi_{0}i_{\kappa}^{\psi_{1}}\exp(\psi_1 (\log K - \log \kappa))-\frac{1}{2}|\sigma_{\kappa}|^{2})v^{(m)}_{\log\kappa}+\frac{|\sigma_{\kappa}|^{2}}{2}v^{(m)}_{\log\kappa,\log\kappa}\\
 & +\mathcal{I}_g(\kappa) \sum_{j=1}^J \pi_g^j [v^{(m,j)}-v^{(m)}].
 \end{split}
\end{equation*}
The first two terms account for the evolution of the knowledge capital and the possibility for R\&D investment to increase the knowledge capital stock. The last term accounts for the possibility of the green technology jump that alters green capital productivity to $A_g^j$ and is increasingly more likely as the knowledge capital stock increases. 

In addition, allowing for R\&D investment alters the output constraint in this setting, so that our flow utility term net of climate damages is now given by 
\begin{align*}
\delta \log ([A_{d}-i_{d}](1-R)+[A_{g}-i_{g}]R - i_{\kappa}).
\end{align*}

We now take first order conditions with respect to $i_d$, $i_g$, and $i_{\kappa}$, which give us expressions for optimal investment choices
\begin{align*}
\delta([A_{d}-i_{d}](1-R)+[A_{g}-i_{g}]R- i_{\kappa})^{-1}	& = \Gamma_{d}\theta_{d}(1+\theta_{d}i_{d})^{-1}[v^{(m)}_{\log K}-Rv^{(m)}_{R}], \\
\delta([A_{d}-i_{d}](1-R)+[A_{g}-i_{g}]R- i_{\kappa})^{-1}	 & = \Gamma_{g}\theta_{g}(1+\theta_{g}i_{g})^{-1}[v^{(m)}_{\log K}+(1-R)v^{(m)}_{R}], \\
\delta([A_{d}-i_{d}](1-R)+[A_{g}-i_{g}]R- i_{\kappa})^{-1}	 & = \psi_{0}\psi_{1}i_{\kappa}^{\psi_{1}-1} \exp(\psi_1 (\log K - \log \kappa)) v^{(m)}_{\log\kappa}. 
\end{align*}

As before, these expressions equate the marginal benefit of an additional unit of a given type of capital to the marginal utility of consumption, which is the utility loss of forgoing consumption for an additional unit of investment. However, given the altered output constraint expression, the marginal utility of consumption on the left-hand side of these equations now also incorporates the fact that some output is dedicated to R\&D investment instead of consumption.

\subsection{Pre Damage and Technology Jumps HJB Equation}

Finally, we have the components needed for the value function associated with pre-technology and -damage jumps. We denote this value function by $V(K_{d},K_{g,}Y,\log \kappa, \log N)$, indicating that neither a realization of $\gamma_3^m$ nor the jump to the green capital productivity has occurred. The same analytical simplification can be applied here, and so the value function in this stage is given by
\begin{align*}
V(K_{d},K_{g,}Y,\log \kappa, \log N)=v(\log K,R,Y_{t},\log \kappa)-\log N_{t}.
\end{align*}

The simplified HJB equation incorporates each of the changes applied to the two intermediate jump state cases. In particular, relative to the post-damage and technology jump state HJB equation, we introduce the following additional terms
\begin{equation*}
\begin{split}
& (-\zeta+\psi_{0}i_{\kappa}^{\psi_{1}}\exp(\psi_1 (\log K - \log \kappa))-\frac{1}{2}|\sigma_{\kappa}|^{2})v_{\log\kappa}+\frac{|\sigma_{\kappa}|^{2}}{2}v_{\log\kappa,\log\kappa}\\
  & - \left( \{\gamma_{1}+\gamma_{2}Y )\}\beta_{f}\eta A_{d}(1-R)K+\frac{1}{2}(\gamma_{2} ) |\varsigma|^{2}(\eta A_{d}(1-R)K)^{2} \right)  \\
 & +\mathcal{I}_g(\kappa) \sum_{j=1}^J \pi_g^j [v^{(j)}-v] + \mathcal{I}_d(y) \sum_{m=1}^{M} \pi_d^m (v^{(m)} - v) .
\end{split}
\end{equation*}
The first line accounts for the evolution of the knowledge capital and the possibility for R\&D investment to increase the knowledge capital stock. The second line captures the altered evolution of damages because the damage jump has not yet occurred. The last line accounts for the possibility of the green technology jump that alters green capital productivity to $A_g^j$ and is increasingly more likely as the knowledge capital stock increases, as well as the expectation of damage function jump that becomes more likely as temperature increases. 

We again alter the output constraint in this setting relative to the post-damage and technology jump case due to the possibility of R\&D investment. As a result, the first order conditions with respect to $i_d$, $i_g$, and $i_{\kappa}$ match those of the intermediate jump state before the technology jump and after the damage jump, except that the relevant value function derivatives do not pertain to $v$. As in each case, the FOCs equate the marginal benefit of an additional unit of a given type of capital to the marginal utility of consumption, which is the utility loss of forgoing consumption for an additional unit of investment.

\begin{comment}
where the remaining HJB equation characterizing the simplified value function is given by:
\begin{equation}
\begin{split}
\delta V & =\max_{i_{g},i_{d}}\delta\log([A_{d}-i_{d}](1-R)+[A_{g}-i_{g}]R -i_{\kappa}) + \delta\log K\\
 & + \left((1-R) \left[\alpha_{d}+\Gamma_d \log (1+ \theta_d i_{d}) \right]+R \left[\alpha_{g}+\Gamma_g \log (1+ \theta_g i_{g})\right]-\frac{   \sigma_d^2 (1-R)^2 + \sigma_g^2 R^2 }{2}  \right)v_{\log K}\\
 &+ \frac{  \sigma_d^2 (1-R)^2 + \sigma_g^2 R^2}{2} v_{\log K,  \log K}  \\
 & +\left( \left[\alpha_{g}+\Gamma_g \log (1+ \theta_g i_{g})\right] - R \sigma_g^2 - \left[\alpha_{d}+\Gamma_d \log (1+ \theta_d i_{d})\right] + (1-R) \sigma_d^2
  \right) R (1-R) v_{R}\\
 & + \frac{1}{2} R^2 (1-R)^2 ({ \sigma_g^2 + \sigma_d^2} ) v_{RR} \\
  & + \left[-R(1-R)^2\sigma_d^2 + R^2(1-R)\sigma_g^2\right]v_{\log K, R} \\
 & +\beta_{f}\eta A_{d}(1-R)K v_{y}+\frac{|\varsigma|^{2}(\eta A_{d}(1-R)K)^{2}}{2} v_{yy}\\
 & - \left( \{\gamma_{1}+\gamma_{2}Y \}\beta_{f}\eta A_{d}(1-R)K+\frac{1}{2}\gamma_{2} |\varsigma|^{2}(A_{d}(1-R)K)^{2} \right) \\
 & +(-\zeta+\psi_{0}i_{\kappa}^{\psi_{1}}\exp(\psi_1 (\log K - \log \kappa))-\frac{1}{2}|\sigma_{\kappa}|^{2})v_{\log\kappa}+\frac{|\sigma_{\kappa}|^{2}}{2}v_{\log\kappa,\log\kappa}\\
 & +(\kappa/\varrho)[v(K,R,Y_{t};A_{g}')-v(K,R,Y_{t},\log\kappa;A_{g})] + \mathcal{I}_d(y)  \sum_{m=1}^{M} \pi_d^m (v^{(m)} - v)
\end{split}
\end{equation}

The FOC for investment and R\&D are unchanged from the baseline setting.
\end{comment}

\section{Model Uncertainty}

With the baseline model results in place, we now introduce the various channels of model uncertainty of focus in our analysis. We explore uncertainty aversion in the form of model misspecification and/or smooth ambiguity across the following channels: 

\begin{itemize}
\item Climate Sensitivity to Carbon Emissions
\vspace{0.1cm}
\item Climate Damage Function Severity and Arrival
\vspace{0.1cm}
%\item {\color{red}Dirty Capital Stock and Investment}
%\vspace{0.1cm}
%\item {\color{red}Green Capital Stock and Investment}
%\vspace{0.1cm}
%\item {\color{red}Knowledge Capital Stock and R\&D Investment}
%\vspace{0.1cm}
%\item {\color{orange}``Green'' Technological Change}
\item ``Green'' Technological Change
%\vspace{0.1cm}
\end{itemize}

%\mb{We have not yet included the items in red into our numerical calculations. The item in orange is only incorporated with a single $A_g^j$ value.}

Aversion to these different channels of uncertainty are introduced into our baseline model using the tools of dynamic decision theory. Importantly, the planner in our model is wary of model uncertainty of these different forms and channels, and thus adopts a preference structure to identify potential worst-case models, constrained by a penalization function, and make optimal policy decisions that are robust to these possible outcomes. We outline the different forms of model uncertainty for our framework in what follows, focusing specifically on misspecification concerns about jump and diffusion processes. Further elaboration on alternative settings, including smooth ambiguity aversion, and complete details on the HJB and FOC equations characterizing the model solutions, is provided in the appendix.

%We focus on the pre damage and technology jump case in our equations, but counterparts can easily adapted to incorporate to the other jump state cases where relevant.

\subsection{Jump Misspecification Concerns}

We first consider the setting with misspecification concerns about the damage function and technological change jump channels. Before the jumps occur, the relevant contributions to the HJB equation related to the technological change and damage function jumps are
\begin{equation*}
\begin{split}
 & \mathcal{I}_g(\kappa) \sum_{j=1}^J \pi_g^j (v^{(j)}-v) + \mathcal{I}_d(y)  \sum_{m=1}^{M} \pi_d^m (v^{(m)} - v). 
 \end{split}
\end{equation*}
 
 Assuming a common robustness parameter $\xi$ across each channel, these terms are replaced by the following uncertainty-adjusted contributions
\begin{equation*}
\begin{split}
\min_{g_j, f_m} \hspace{0.1cm} & \mathcal{I}_g(\kappa) \sum_{j=1}^J \pi_g^j g_j (v^{(j)}-v) + \mathcal{I}_d(y)  \sum_{m=1}^{M} \pi_d^m f_m (v^{(m)} - v) \\
 & + \xi \left[ \mathcal{I}_g(\kappa) \sum_{j=1}^J \pi_g^j \left( 1 - g_j + g_j \log g_j \right)  +  \mathcal{I}_d(y) \sum \pi_d^m \left( 1 - f_m + f_m \log f_m \right) \right].
\end{split}
\end{equation*}

The adjustment introduces probability distortions $g_j$ and $f_m$ to the jump processes for technological change and climate damages. These distortions impact expectations about the arrival rate and distribution of post-jump outcomes related to these jumps. To constrain these distortions, the relative entropy terms $\xi \left[ \mathcal{I}_g(\kappa) \sum_{j=1}^J \pi_g^j \left( 1 - g_j + g_j \log g_j \right)  +  \mathcal{I}_d(y) \sum \pi_d^m \left( 1 - f_m + f_m \log f_m \right) \right]$ are introduced into preferences, penalizing distortions that are ``too large'' based on the uncertainty aversion parameter $\xi$. Note that $g_j$, and $f_m$ are endogenous objects solved for by the decision maker by minimizing discounted lifetime expected utility, subject to the relative entropy penalties. The FOCs resulting from this minimization objective are given by
\begin{eqnarray*}
g_j & = & \exp \left(- \frac{1}{\xi} (v^{(j)}-v) \right), \cr
f_m & = & \exp \left( -\frac{1}{\xi} (v^{(m)} - v) \right). 
\end{eqnarray*}

Each of these represents probability distortions that adjust the distributions associated with the jump processes related to the economic and climate channels for which there are concerns about model uncertainty. The inclusion of these probability distortions into the planner's problem leads to robustly-altered optimal policy choices that account for the distorted likelihood of outcomes based on potential worst-case outcomes. 

We note that in each of the various jump realization states, there are modifications that need to be made to these components. After the technology and damage jumps have both taken place, jump uncertainty is no longer relevant for the social planner. After the technology jump has occurred, but not the damage jump, we replace $A_g$ with $A_g^j$ and the relevant uncertainty adjustments to the HJB equation and minimization FOC pertaining to the climate model and damage function jump are given by
\begin{equation*}
\begin{split}
\min_{f_m} \hspace{0.1cm} & \mathcal{I}_d(y)  \sum_{m=1}^{M} \pi_d^m f_m (v^{(m,j)} - v^{(j)})  + \xi \mathcal{I}_d(y) \sum \pi_d^m \left( 1 - f_m + f_m \log f_m \right), \\
f_m & =  \exp \left( -\frac{1}{\xi} (v^{(m,j)} - v^{(j)}) \right). \cr
\end{split}
\end{equation*}

Before the technology jump has occurred, but after the damage jump, the relevant uncertainty adjustments to the HJB equation and minimization FOC pertaining to the climate model and technology jump are as follows
\begin{equation*}
\begin{split}
\min_{g_j} \hspace{0.1cm} & \mathcal{I}_g(\kappa) \sum_{j=1}^J \pi_g^j g_j (v^{(m,j)}-v^{(m)}) + \xi \mathcal{I}_g(\kappa)  \sum_{j=1}^J \pi_g^j \left( 1 - g_j + g_j \log g_j \right), \\
 g_j & = \exp \left(- \frac{1}{\xi} (v^{(m,j)}-v^{(m)}) \right).
\end{split}
\end{equation*}

\subsection{Diffusion Misspecification Concerns}

We now consider the setting with misspecification concerns about the climate and capital stocks diffusion channels. Before the technology and damage jumps occur, the relevant contributions to the HJB equation related to the diffusion components are of the form
\begin{equation*}
\begin{split}
 & \left((1-R) \left[\alpha_{d}+\Gamma_d \log (1+ \theta_d i_{d}) \right]+R \left[\alpha_{g}+\Gamma_g \log (1+ \theta_g i_{g}) \right] -\frac{   \sigma_d^2 (1-R)^2 + \sigma_g^2 R^2 }{2} \right)v_{\log K} \\
 & +\left( \left[\alpha_{g}+\Gamma_g \log (1+ \theta_g i_{g})\right] - R \sigma_g^2 - \left[\alpha_{d}+\Gamma_d \log (1+ \theta_d i_{d})\right] + (1-R) \sigma_d^2  \right) R (1-R) v_{R}\\
 & + \left( \left[ v_{y} - \{\gamma_{1}+\gamma_{2}Y \} \right]\beta_{f} \eta A_{d}(1-R)K+\frac{1}{2}\gamma_{2} |\varsigma|^{2}(\eta A_{d}(1-R)K)^{2} \right) \\
 & +\left(-\zeta+\psi_{0}i_{\kappa}^{\psi_{1}}\exp(\psi_1 (\log K - \log \kappa) )-\frac{1}{2}|\sigma_{\kappa}|^{2}  \right) v_{\log \kappa}.  \\
\end{split}
\end{equation*}

 Assuming the same common robustness parameter $\xi$ for each channel as before, these terms are replaced by the following uncertainty-adjusted contributions
\begin{equation*}
\begin{split}
\min_{h} \hspace{0.1cm} & \left((1-R) \left[\alpha_{d}+\Gamma_d \log (1+ \theta_d i_{d}) \right]+R \left[\alpha_{g}+\Gamma_g \log (1+ \theta_g i_{g}) \right] -\frac{   \sigma_d^2 (1-R)^2 + \sigma_g^2 R^2 }{2} \right)v_{\log K} \\
 & +\left( \left[\alpha_{g}+\Gamma_g \log (1+ \theta_g i_{g})\right] - R \sigma_g^2 - \left[\alpha_{d}+\Gamma_d \log (1+ \theta_d i_{d})\right] + (1-R) \sigma_d^2  \right) R (1-R) v_{R}\\
 & + \left( \left( v_{\log K} -  R v_{R} \right) (1-R) \sigma_d  + \left( v_{\log K} + (1-R) v_{R} \right)  R \sigma_g \right) \cdot h \\
 & + \left( \left[ v_{y} - \{\gamma_{1}+\gamma_{2}Y \} \right](\beta_{f} + \varsigma \cdot h)\eta A_{d}(1-R)K+\frac{1}{2}\gamma_{2} |\varsigma|^{2}(\eta A_{d}(1-R)K)^{2} \right) \\
 & +\left(-\zeta+\psi_{0}i_{\kappa}^{\psi_{1}}\exp(\psi_1 (\log K - \log \kappa) )-\frac{1}{2}|\sigma_{\kappa}|^{2} + \sigma_{\kappa} \cdot h \right) v_{\log \kappa} + \xi \frac{|h|^2}{2}. 
\end{split}
\end{equation*}

This adjustment introduces a drift distortion $h$ to the diffusion processes for the different types of capitals and for temperature, disguised by the Brownian shocks. This type of drift distortion is a known result from the Girsanov theorem, and captures a distributional change represented by a likelihood ratio. To constrain this distortion, again, a relative entropy term is introduced into preferences, in this setting, the quadratic expression $\xi |h|^2/2$, penalizing distortions that are ``too large'' based on the uncertainty aversion parameter $\xi$. As with $g_j$ and $f_m$, $h$ is an endogenous object solved by the decision maker by minimizing discounted lifetime expected utility, subject to the relative entropy penalties. The FOC resulting from this minimization objective is given by
\begin{eqnarray*}
h   & = &  - \frac{1}{\xi} \left[ \left( v_{\log K} -  R v_{R} \right) (1-R) \sigma_d + \left( v_{\log K} + (1-R) v_{R} \right)  R \sigma_g \right] \\
    &   & - \frac{1}{\xi} \varsigma \eta A_{d}(1-R)K \left(v_{y} -  \{\gamma_{1}+\gamma_{2}Y \}\right) \\
    &   & - \frac{1}{\xi} \sigma_{\kappa} v_{\log \kappa}. 
\end{eqnarray*}

The inclusion into the planner's problem of this probability distortion in the form of a drift distortion leads to further robustly-altered optimal policy choices that account for the distorted likelihood of outcomes based on potential worst-case outcomes. 

%There are explicit contributions for each of the capital stocks and for temperature related to diffusion uncertainty in the model. However, allowing for all of the uncertainty channels simultaneously means that uncertainty contributions will have interaction effects as the jump misspecification distortions $f_m$ and $g_j$, and the diffusion misspecification distortion $h$, will alter the value functions, and their derivatives, across jump states and states of nature, influencing the optimized distortions arising from uncertainty aversion and the optimal policy choices made by the social planner.

As with the jump misspecification concerns, in each of the various jump realization states there are modifications that need to be made to these components. After the technology and damage jumps have both taken place, the remaining uncertainty implications come from the climate sensitivity channel and the green and dirty capital stocks. Therefore, in addition to replacing $A_g$ with $A_g^j$, and denoting the realized $\gamma_3^m$ value, the uncertainty-adjusted contribution to the HJB and the relevant minimization FOC are given by
\begin{equation*}
\begin{split}
\min_{h} \hspace{0.1cm} & \left((1-R) \left[\alpha_{d}+\Gamma_d \log (1+ \theta_d i_{d}) \right]+R \left[\alpha_{g}+\Gamma_g \log (1+ \theta_g i_{g}) \right] -\frac{   \sigma_d^2 (1-R)^2 + \sigma_g^2 R^2 }{2} \right)v^{(m,j)}_{\log K} \\
 & +\left( \left[\alpha_{g}+\Gamma_g \log (1+ \theta_g i_{g})\right] - R \sigma_g^2 - \left[\alpha_{d}+\Gamma_d \log (1+ \theta_d i_{d})\right] + (1-R) \sigma_d^2  \right) R (1-R) v^{(m,j)}_{R}\\
 & + \left( \left( v^{(m,j)}_{\log K} -  R v^{(m,j)}_{R} \right) (1-R) \sigma_d  + \left( v^{(m,j)}_{\log K} + (1-R) v^{(m,j)}_{R} \right)  R \sigma_g \right) \cdot h \\
 & + \left( \left[ v^{(m,j)}_{y} - \{\gamma_{1}+\gamma_{2} \hat{Y} + \gamma_3^m(\hat{Y} - \tilde{y}) \} \right](\beta_{f} + \varsigma \cdot h)\eta A_{d}(1-R)K \right) + \xi \frac{|h|^2}{2}, \\
 h   & =  - \frac{1}{\xi} \left[ \left( v^{(m,j)}_{\log K} -  R v^{(m,j)}_{R} \right) (1-R) \sigma_d + \left( v^{(m,j)}_{\log K} + (1-R) v^{(m,j)}_{R} \right)  R \sigma_g \right] \\
    &  - \frac{1}{\xi} \varsigma \eta A_{d}(1-R)K \left(v^{(m,j)}_{y} -  \{\gamma_{1}+\gamma_{2}\hat{Y} + \gamma_3^m(\hat{Y} - \tilde{y}) \}\right).
\end{split}
\end{equation*}

After the technology jump has occurred, but not the damage jump, we replace $A_g$ with $A_g^j$ and the relevant uncertainty adjustments to the HJB equation and minimization FOC pertaining to the diffusion terms are given by
\begin{equation*}
\begin{split}
\min_{h} \hspace{0.1cm} & \left((1-R) \left[\alpha_{d}+\Gamma_d \log (1+ \theta_d i_{d}) \right]+R \left[\alpha_{g}+\Gamma_g \log (1+ \theta_g i_{g}) \right] -\frac{   \sigma_d^2 (1-R)^2 + \sigma_g^2 R^2 }{2} \right)v^{(j)}_{\log K} \\
 & +\left( \left[\alpha_{g}+\Gamma_g \log (1+ \theta_g i_{g})\right] - R \sigma_g^2 - \left[\alpha_{d}+\Gamma_d \log (1+ \theta_d i_{d})\right] + (1-R) \sigma_d^2  \right) R (1-R) v^{(j)}_{R}\\
 & + \left( \left( v^{(j)}_{\log K} -  R v^{(j)}_{R} \right) (1-R) \sigma_d  + \left( v^{(j)}_{\log K} + (1-R) v^{(j)}_{R} \right)  R \sigma_g \right) \cdot h \\
 & + \left( \left[ v^{(j)}_{y} - \{\gamma_{1}+\gamma_{2}Y  \} \right](\beta_{f} + \varsigma \cdot h)\eta A_{d}(1-R)K \right) + \xi \frac{|h|^2}{2}, \\
 h   & =  - \frac{1}{\xi} \left[ \left( v^{(j)}_{\log K} -  R v^{(j)}_{R} \right) (1-R) \sigma_d + \left( v^{(j)}_{\log K} + (1-R) v^{(j)}_{R} \right)  R \sigma_g \right] \\
    &  - \frac{1}{\xi} \varsigma \eta A_{d}(1-R)K \left(v^{(j)}_{y} -  \{\gamma_{1}+\gamma_{2}Y \}\right).
\end{split}
\end{equation*}

Before the technology jump has occurred, but after the damage jump, the relevant uncertainty adjustments to the HJB equation and minimization FOC pertaining to the diffusion terms are
\begin{equation*}
\begin{split}
\min_{h} \hspace{0.1cm} & \left((1-R) \left[\alpha_{d}+\Gamma_d \log (1+ \theta_d i_{d}) \right]+R \left[\alpha_{g}+\Gamma_g \log (1+ \theta_g i_{g}) \right] -\frac{   \sigma_d^2 (1-R)^2 + \sigma_g^2 R^2 }{2} \right)v^{(m)}_{\log K} \\
 & +\left( \left[\alpha_{g}+\Gamma_g \log (1+ \theta_g i_{g})\right] - R \sigma_g^2 - \left[\alpha_{d}+\Gamma_d \log (1+ \theta_d i_{d})\right] + (1-R) \sigma_d^2  \right) R (1-R) v^{(m)}_{R}\\
 & + \left( \left( v^{(m)}_{\log K} -  R v^{(m)}_{R} \right) (1-R) \sigma_d  + \left( v^{(m)}_{\log K} + (1-R) v^{(m)}_{R} \right)  R \sigma_g \right) \cdot h \\
 & + \left( \left[ v^{(m)}_{y} - \{\gamma_{1}+\gamma_{2}\hat{Y} + \gamma_3^m(\hat{Y} - \tilde{y}) \} \right](\beta_{f} + \varsigma \cdot h)\eta A_{d}(1-R)K+\frac{1}{2}\gamma_{2} |\varsigma|^{2}(\eta A_{d}(1-R)K)^{2} \right) \\
 & +\left(-\zeta+\psi_{0}i_{\kappa}^{\psi_{1}}\exp(\psi_1 (\log K - \log \kappa) )-\frac{1}{2}|\sigma_{\kappa}|^{2} + \sigma_{\kappa} \cdot h \right) v^{(m)}_{\log \kappa} + \xi \frac{|h|^2}{2}, \\
 h  & = - \frac{1}{\xi} \left[ \left( v^{(m)}_{\log K} -  R v^{(m)}_{R} \right) (1-R) \sigma_d + \left( v^{(m)}_{\log K} + (1-R) v^{(m)}_{R} \right)  R \sigma_g \right] \\
    & - \frac{1}{\xi} \varsigma \eta A_{d}(1-R)K \left(v^{(m)}_{y} -  \{\gamma_{1}+\gamma_{2}\hat{Y} + \gamma_3^m(\hat{Y} - \tilde{y}) \}\right) \\
    & - \frac{1}{\xi} \sigma_{\kappa} v^{(m)}_{\log \kappa}.
\end{split}
\end{equation*}

\subsection{Full Misspecification Concerns}

We have so far shown the introduction of the jump and diffusion uncertainty separately in our analysis. However, in our analysis, we incorporate each of the uncertainty channels simultaneously, allowing for broadly conceived uncertainty considerations in our results. There are explicit contributions related to diffusion uncertainty and jump uncertainty in the model. However, allowing for all of the uncertainty channels simultaneously allows for important interaction effects as the jump misspecification distortions $f_m$ and $g_j$, and the diffusion misspecification distortion $h$, will alter the value functions and their derivatives across jump states and states of nature. These more implicit effects will influence the optimized distortions arising from uncertainty aversion, and as a result, impact the social valuations and optimal policy choices of the social planner.

\section{Computational Method}

Before delving into numerical results, we outline the numerical method used to derive those results. The computational algorithm developed here is a critical contribution of our paper to the literature. Much of the theoretical work in climate economics and climate finance requires, by necessity, the use of computational methods to derive solutions. The integrated structure of these models requires a multi-dimensional state space, often times with non-linear dynamic relationships and functional forms, in order to rigorously account for relevant model features. The result of specifying such models is the need to confront the ``curse of dimensionality'', or the fact that the complexity of deriving numerical solutions exponentially increases in the number of dimensions. As noted by \citep{HaJeE:18}, numerical solutions are almost always unavailable for problems where the number of dimensions is greater than or equal to 4 when solving with standard finite difference or finite element methods. Incorporating model uncertainty, which introduces non-linear endogenous responses to model uncertainty, such as exponential tilting to the distribution of expected future model outcomes, can further exacerbate the computational burden. As we will outline below, we provide a novel implementation of deep neural networks for deriving the numerical solutions to our HJB equations. 

Only very recently have deep learning and neural network methods been applied to solving dynamic economic models. Because of the ability of these methods to provide global solutions for problems with high dimensions where the ``curse of dimensionality'' can make computational solutions infeasible for other methods, or where significant non-linearities can cause other computational methodologies to fail, deep learning solution methods have been seen as a potentially ``game-changing'' toolset. By approximating the representative agent's value function and optimal controls with deep neural networks, the problems of exponentially increasing complexity from high-dimensional state spaces can be ameliorated due to the representation of functions in a compositional form, rather than by the standard additive form resulting from finite difference and element methods. Applications in macroeconomics \citep{fernandez2020solving, maliar2021deep, azinovic2022deep} and finance \citep{duarte2018machine, sauzet2021projection} implementing these types of solutions methods are only beginning to scratch the surface of the potential value and importance for study key problems in the literature more broadly.

% {\color{blue}
While this existing work has explored settings with (potentially many) more state variables than our current setting, they tend to rely on symmetric, stationary, and exogenous state variables, as well as linearity in the model framework, to maintain tractability with such scale. Our methodology allows us to go beyond these settings, which is essential to analyzing our climate-economics model with model uncertainty aversion. Specifically, while an improvised finite difference method using neural networks to solve linear systems may help mitigate the curse of dimensionality, it does not produce satisfactory results for complex models with strong nonlinearities such as ours. Also, due to the strong nonlinearities in our model, naively parameterizing the value function using a neural network and minimizing the loss associated with the PDE operator (like the standard deep Galerkin method) will not produce accurate enough results. While an extended version of the deep Galerkin method \citep{al2022extensions} is indeed helpful in improving accuracy, it does not effectively preserve the value function's monotonicity with respect to certain parameters which are supported by economic arguments. 

We address these issues in our framework by implementing an extended deep Galerkin method algorithm that considers critical parameters as additional (pseudo) variables in the network input, an approach not previously explored in the literature. Because of this, we are able to derive global numerical solutions for a continuous-time, infinite horizon setting with significant non-linearities and multiple endogenous state variables. This is critical to our analysis, as exploring transition dynamics for endogenous optimal policy responses is at the heart of understanding and analyzing outcomes related to climate change, model uncertainty, and the transition to carbon-neutrality. Furthermore, our algorithm can still handle high-dimensional PDEs, avoiding the ``curse of dimensionality'' in such cases. As such, the resiliency of our algorithm to various modeling complexities, while still being able to scale to higher-dimensional settings, opens the door to exploring models with regional, household, and firm heterogeneities across technologies, economic frictions, policy objectives, and other sources. Thus, our numerical algorithm not only enriches our ability to explore a more extensive set of questions and models in climate-economics and climate finance than before, but also more broadly in economics, finance, and other areas of research that require solving dynamic stochastic optimal control problems.

\subsection{Implementing Neural Nets for Numerical Solutions}\label{sec:DGM-PIA}

We now outline our algorithm, and provide pseudo-code and further details about implementation in the appendix. We use the deep Galerkin method-policy improvement algorithms (DGM-PIA) proposed in \cite{al2022extensions} to solve the aforementioned HJB equations. We first illustrate the algorithm on a generic HJB equation for $V(\bm x)$:
\begin{equation*}\label{eq:HJB_generic}
    - \delta V(\bm x) + \sup_{\bm \alpha \in \mc{A}}\{\mc{L}^{\bm \alpha} V( \bm x) + f(\bm x, \bm \alpha) \} = 0,
\end{equation*}
where $\bm x$ and $\bm \alpha$ denote the state and control variables, $\mc{A}$ is the control space, the differential operator $\mc{L}^{\bm \alpha}$ is the infinitesimal generator of the controlled state process $\bm X^{\bm \alpha}$,   $f$ is the utility function and $\delta$ is the discount factor. DGM-PIA solves for the value function $V$ and the optimal control $\bm \alpha$ simultaneously by parameterizing both as deep neural networks $V^\theta$ and $\bm \alpha^\varphi$. Then, the networks are trained by taking alternating stochastic gradient descent steps for the two functions. Let $\bm \alpha^{\varphi_0}$ (as a function of $\bm x$) be the initial control parameterized by the neural net,  at stage $n$, the algorithm contains two steps:

\vspace{0.25cm}

\noindent Step 1. Find a solution to the linear PDE
\begin{equation*}
   - \delta V^{\theta_n}(\bm x) + \mc{L}^{\bm \alpha^{\varphi_{n}}}V^{\theta_n}(\bm x) + f(\bm x, \bm\alpha^{\varphi_{n}}(\bm x)) = 0,
\end{equation*}
for the fixed control $\bm \alpha^{\varphi_n}$, by updating $\theta_n$ via minimizing
\begin{equation*}\label{eq:DGM_HJB}
    L_V (\theta ) = \lVert  - \delta V^{\theta}(\bm x) + \mc{L}^{\bm \alpha^{\varphi_n}}V^{\theta}(\bm x) + f(\bm x, \bm\alpha^{\varphi_n}(\bm x))  \rVert^2.
\end{equation*}

\noindent Step 2. Update the policy corresponding to 
\begin{equation*}
\bm \alpha^{\varphi_{n+1}}(\bm x) \in \argmax_{\bm \alpha \in \mc{A}} \{ \mc{L}^{\bm \alpha} V^{\theta_n}(\bm x) + f(\bm x, \bm \alpha)\},
\end{equation*}
for the fixed value function $V^{\theta_n}$, by update $\varphi_{n+1}$ via minimizing
\begin{equation*}\label{eq:DGM_ctrl}
    L_{\bm \alpha}(\varphi) = - \int_\Omega \left[ \mc{L}^{\bm \alpha^\varphi} V^{\theta_n}(\bm x) + f(\bm x, \bm \alpha^\varphi(\bm x)) \right] \ud \nu(\bm x),
\end{equation*}
where $\nu(\bm x)$ is a probability measure on the domain $\Omega$ of $\bm x$ characterizing the different regions' relative importance. 

\vspace{0.25cm}

%In our problem, depending on the setting we are solving, the state processes could contain $(K_d, K_g, Y_t, \log \mc{I}_g, \log N_t)$ \rh{$(\log K, R, Y_t, \log \kappa, \log N_t)$?},  the control variables could contain $(i_d, i_g, i_{\mc{I}}, g, f_m, h, \omega_\ell, g_j)$ \rh{$(i_d, i_g, i_{\kappa}, g_j, f_m, h)$?}, $\mc{L}^{\bm \alpha}$ is the corresponding infinitesimal generator.

In our problem, depending on the setting we are solving, the state processes could contain $(\log K, R, Y_t, \log \kappa, \log N_t)$, the control variables could contain $(i_d, i_g, i_{\kappa}, g_j, f_m, h)$. In addition, in order to keep the value function's monotonicity with respect to $\gamma_3^m$, we take $\gamma_3^m$ as an input or ``pseudo-state'' of the parameterized neural network.

%Do we want to mention that, in order to keep the value function's monotonicity w.r.t. $\gamma_3$, we take $\gamma_3$ as an input of the parameterized neural network?

\section{Numerical Results}

We next present and discuss the numerical model solution results, which are derived using the numerical algorithm outlined above. Before getting into the results, we briefly outline a few details for completeness regarding model assumptions, functional forms, and parameter values. After presenting and discussing the numerical results, we discuss details about how we validate our neural-network-based solutions.

\subsection{Functional Forms and Assumptions}

First, for tractability we consider the case of independent Brownian shocks, i.e.,
\begin{comment}
\begin{align*}
\sigma_{d}' & =[\bar{\sigma}_{d},0,0,0]\\
\sigma_{g}' & =[0,\bar{\sigma}_{g},0,0]\\
\varsigma' & =[0,0,\bar{\varsigma},0]\\
\sigma_{\kappa}' & =[0,0,0,\bar{\sigma}_{\kappa}]
\end{align*}
such that the following holds
\end{comment}
\begin{align*}
0 & =\sigma_{d}'\sigma_{g}=\sigma_{d}'\varsigma=\sigma_{d}'\sigma_{\kappa},\\
0 & =\sigma_{g}'\varsigma=\sigma_{g}'\sigma_{\kappa},\\
0 & =\varsigma'\sigma_{\kappa}.
\end{align*}

Second, we use the following functional forms for our jump arrival rates. For the technology change jump, we assume an arrival rate that is proportional to the knowledge stock: $\mathcal{I}_g(\kappa ) = \kappa /\varrho$. The parameter $\varrho$ scales the knowledge capital stock variable to change the units into arrival rate units, and is chosen based on expected green policy implementation timelines proposed by various countries. Section \ref{section:parameters} provides further details about the choice of this parameter value.

The damage jump intensity $\mathcal{I}_d(y)$ follows the arrival rate proposed in \cite{BarnettBrockHansen:2021}:
\begin{eqnarray*}
    \mathcal{I}_d(y) = r_1 \left( \exp \left[ \frac{r_2}{2}(y- \underline{y})^2 \right] - 1 \right) \mathbbm{1}_{y \ge \underline{y}}.
\end{eqnarray*}

\begin{figure}[!pht]
	\centering
	\includegraphics[width=0.75\textwidth]{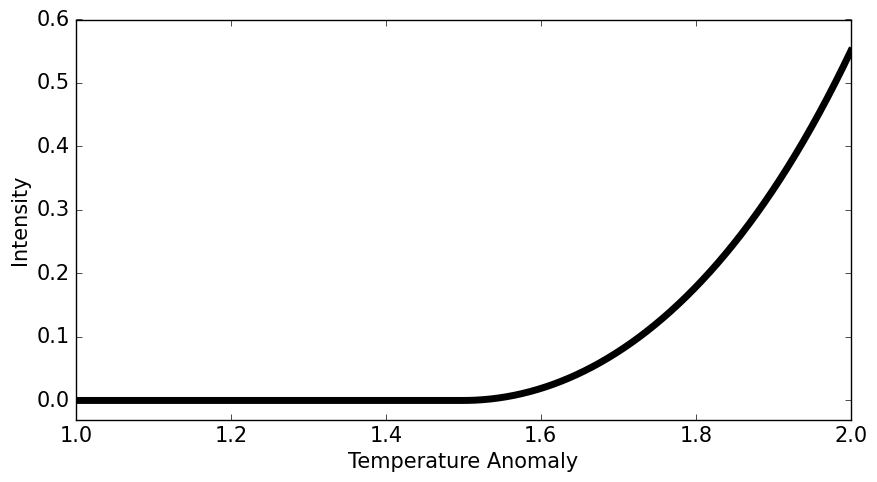}
\caption{Intensity function, $r_1=1.5$ and $r_2=2.5$. With this intensity function, the probability of a jump at an anomaly of $1.6$ is approximately $.02$ per annum, increasing to about $.08$ per annum
at an anomaly of $1.7$, increasing further to approximately $.18$ per annum at an anomaly of $1.8$ and then to about one third per annum when the anomaly is $1.9$.}\label{fig:damage_arrival_rate}
%\parbox{\textwidth}{\footnotesize Intensity function, $r_1=1.5$ and $r_2=2.5$. With this intensity function, the probability of a jump at an anomaly of $1.6$ is approximately $.02$ per annum, increasing to about $.08$ per annum at an anomaly of $1.7$, increasing further to approximately $.18$ per annum at an anomaly of $1.8$ and then to about one third per annum when the anomaly is $1.9$.}
\end{figure}

Figure \ref{fig:damage_arrival_rate} shows the increasing nature of the arrival rate as temperature anomaly $y$ increases. The calibration is such that the probability of a jump taking place by $y = 2$ is essentially 1. 

\subsection{Parameter Values} \label{section:parameters}

We now outline how we chose the parameter values used in our numerical analysis. The economic parameter values are given in Table \ref{table:econ_parameters}. We note that the special case of our model without climate change and technological innovation is the model given in \cite{EberlyWang:2009}. We therefore use their assumed parameters values of the relevant economic parameters as our baseline solution values. These parameters are not without justification from empirical estimates and previous modeling set-ups. The choice of $\delta$ is consistent with values of the subjective discount rate used in the macroeconomics and asset pricing literature. $(\Gamma_d, \theta_d)$ produces a value of Tobin's q of 2.5, which is consistent with estimated values in the macroeconomics literature. The values of $\sigma_d$ and $\sigma_g$ are within the range of values used in the macro-asset pricing literature to produce values of the price of risk observed in the data. The values of $A_d$ and $A_g$ generate output consistent with World Bank World GDP values. \cite{way2022empirically} provide estimates of future costs of green technology across various scenarios (fast transition, slow transition, no transition) by estimating coefficients for stochastic Wright's law and Moore's law. The values for $A_g^j$ coincide with implied productivity gains in different green technologies over the next 50 years for these estimates.

\begin{table}[H]
\caption{Economic Parameters}\label{table:econ_parameters}
	\centering
	\begin{tabular}{ll}
   \hline
		Parameters & values\\
		\toprule
		$\delta$ & 0.025\\
		$(\alpha_d, \Gamma_d, \theta_d, \sigma_d)$ & ( -0.035, 0.025, 100, 0.15)\\
		$(\alpha_g, \Gamma_g, \theta_g, \sigma_g)$ & ( -0.035, 0.025, 100, 0.15) \\
		$(A_d, A_g)$ & (0.12, 0.10) \\
		$ \{A_g^j\} $ & $ A_g \times \{1 + \frac{(j-1)}{J-1} \}_{j = 1, \dots, 3}$ \\
		$(\zeta, \psi_0, \psi_1, \sigma_{\kappa})$ & (0, 0.10583, 0.5, 0.016)\\
        $\varrho$ & 448 \\%{\color{red} (Alt: 1120)}\\
		\bottomrule
	\end{tabular}
 
 %{\footnotesize {\color{red}Alternative parameterization in appendix.}}
 
\end{table}

%$\{0.15, 0.20, 0.30\} $ \\
%given by
%\begin{eqnarray*}
%A_g^j \in \{  A_g \times(1 + \frac{(j-1)}{J-1}) \}_{j = 1, \dots, J}
%\end{eqnarray*} 

We assume $\zeta = 0$. \cite{lucking2019have} and \cite{bloom2019toolkit} have provided estimates for the returns to R\&D investment which guide our choice of $(\psi_0, \psi_1)$. The choice of $\varrho$ translates our initial value of knowledge stock, which is based on estimates from the BLS of the total US R\&D stock values (scaled up to World values), to an expected arrival time of a green technological innovation occurring between 30 and 80 years. The value of $\sigma_{\kappa}$ is chosen to match the other capital volatilities, which are based on estimates from the World Bank database.

\begin{table}[H]
\caption{Climate Dynamics and Damages Parameters} \label{table:climate_parameters}
	\centering
	\begin{tabular}{ll}
   \hline
		Parameters & values\\
		\toprule
		$\beta_f$ & 1.86 / 1000\\
        $\eta$ & 0.17 \\
		$\varsigma$ & $1.2 \times 1.86 / 1000$ \\
		$(\gamma_1, \gamma_2)$ &  $(0.00017675, 2 \times 0.0022)$ \\
		$\{ \gamma_3^m \}$ & $\{\frac{1}{3} \frac{m-1}{M-1}\}_{m = 1, \dots, 5}$\\
        $(r_1, r_2, \underline{y})$ & $(1.5, 2.5, 1.5)$ \\
		$\bar{y}$ & 2\\
		\bottomrule
	\end{tabular}
\end{table}

The climate dynamics and climate damage parameter values are given in Table \ref{table:climate_parameters}. The values of $\beta_{f,\ell}$ come from pulse experiments estimates produced by \cite{BarnettBrockHansen:2021}, based on the results from \cite{Joosetal:2013} and \cite{Geoffroy:2013}, and are consistent with values reported in \cite{IPCC:2021}. The value $\beta_{f}$ is the average value across all climate 144 models. The value of $\eta$ is chosen to match the current estimate of annual carbon emissions of 10 GtC from \cite{Figueresetal:2018}, based on World Bank estimates of World GDP and EIA/IEA estimates of the clean and dirty capital split. The value of $\varsigma$ matches volatility used by \cite{BarnettBrockHansen:2021}. The values of $\gamma_1$, $\gamma_2$, $\gamma_3^m$, and $\bar{y}$ match damage function parameters used by \cite{BarnettBrockHansen:2021}, which are designed to incorporate the spread of potential climate damage outcomes based on \cite{Nordhaus:2019}, \cite{Weitzman:2012}, and others in the literature.

\begin{table}[H]
\caption{State Variable Initial Values and Ranges}\label{table:state_space}
	\centering
	\begin{tabular}{ll}
   \hline
		State variables & values \\
		\toprule
		$K_0$ & 739 \\
        $R_0$ & $0.5$ \\
        $Y_0$ & 1.1 \\
		$\kappa_{0}$ &  $11.2$\\
%	  $\varrho$ &   $448$ \\
  \hline
		State variables & range\\  
    \hline
   $\log(K)$ & $ [4, 8.5]$ \\
	$R $ & $[0.01, 0.99]$ \\
	$Y $ & $[0, 4]$ \\
	$\log(\kappa) $ & $[1 , 6]$ \\
		\bottomrule
	\end{tabular}
\end{table}

For our computations, we must also specify ranges and initial values for our state variables. These values are given in Table \ref{table:state_space}. The initial value of total capital $K_0$ matches estimates from the World Bank of World GDP. The initial value of global mean temperature anomaly $Y_0$ matches the estimated current value from \cite{IPCC:2021}. The initial value of the green capital-to-total capital ratio $R_0$ is based on estimates of clean and dirty capital splits from the EIA and IEA. The initial value of knowledge stock $\kappa_{0}$ matches estimates from the BLS of Total US R\&D stock values (scaled up to World values).

\begin{comment}
Matching log adjustment costs and quadratic adjustment costs:

Define the equation
\begin{align*}
f(i) & = \alpha+\Gamma\log(1+\theta i)
\end{align*}

The 2nd order Taylor expansion for this around $\bar{i}$ is given by
\begin{align*}
\tilde{f}(i) & = \alpha+\Gamma\log(1+\theta\bar{i})+\Gamma\theta(1+\theta\bar{i})^{-1}(i-\bar{i})-\frac{\Gamma\theta^{2}}{2}(1+\theta\bar{i})^{-2}(i-\bar{i})^{2}
\end{align*}

At $\bar{i}=0$, we simplify our expression as follows
\begin{align*}
\tilde{f}(i) & = \alpha+\Gamma\theta i-\frac{\Gamma\theta^{2}}{2}(i)^{2}
\end{align*}

Defining the quadratic adjustment cost function as $g(i) = \alpha+i-\frac{\phi}{2}(\theta^2)$, then we can match up coefficients to get:
\begin{align*}
\alpha & = \alpha \\
\Gamma & = \frac{1}{\phi} \\
\theta & = \frac{1}{\Gamma} = \phi
\end{align*}

State space range:

\begin{align*}
	\log(K) \in & [4, 8.5]\\
	R \in & [0.01, 0.99]\\
	Y \in& [0, 4]\\
	\log(\kappa) \in& [1 , 6]
\end{align*}

\end{comment}

\subsection{Model Solution Results}

We now discuss the computational results of our model. The results shown are simulation pathway outcomes based on the solutions to the HJB equations and are initialized at today's values of the state variables and shown out to 30 years, which is near the time when the temperature anomaly hits $1.5^{\circ} C$ and the probability of a damage jump occurring becomes non-zero. For the probability of the technology and damage jumps occurring, we examine the outcomes out to 40 years. The results provided are for the following specifications related to model uncertainty:
%For the probability distribution of climate and damage models, which are adjusted for uncertainty concerns when $\xi < \infty$, and the jump probabilities, we examine the outcomes out to 30 and 40 years.
%\vspace{0.5cm}
\begin{itemize}
%\item 144 climate models and  5 damage models 
\item $\beta_f$ as the average of the 144 climate models 
\item 20 damage models $\gamma_3 \in \{0, ... , 1/3\}$
\item Misspecification over technology and damage jumps and the climate model dynamics
%\item Comparative statics across post-jump green technology $A_g' \in \{ 0.15, 0.3 \}$
\item Comparison across uncertainty parameters $\xi \in \{ 0.1, \infty \}$ 
\end{itemize}

\begin{figure}[!pht]

%\vspace{-1.5cm}
%\vspace{-0.5cm}

%\caption{Numerical Results - Emissions, R\&D, and Investment}  \label{fig:ems_rd_inv}
\begin{center}

        \begin{subfigure}[b]{0.495\textwidth}  
            \centering 
            \includegraphics[width=\textwidth]{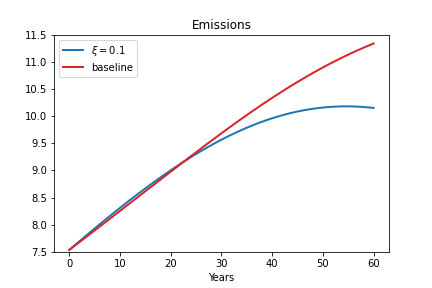}
            \caption[Network2]%
            {{\small Gigatons of carbon emissions}}\label{fig:PreJumpEms}           
        \end{subfigure}
        \hfill
        \begin{subfigure}[b]{0.495\textwidth}
            \centering
            \includegraphics[width=\textwidth]{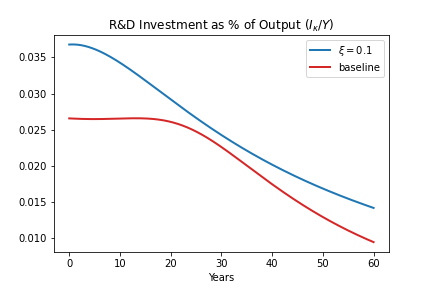}
            \caption[]%
            {{\small R\&D investment-to-output ratio}}\label{fig:PreJumpRD}                 
        \end{subfigure}
        
        \vskip\baselineskip
        
        \begin{subfigure}[b]{0.495\textwidth}   
            \centering 
            \includegraphics[width=\textwidth]{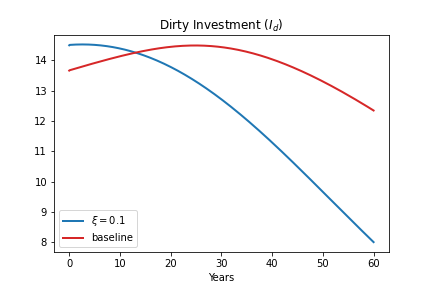}
            \caption[]%
            {{\small Dirty capital investment}}\label{fig:PreJumpId}    
        \end{subfigure}
                \hfill
        \begin{subfigure}[b]{0.495\textwidth}
            \centering
            \includegraphics[width=\textwidth]{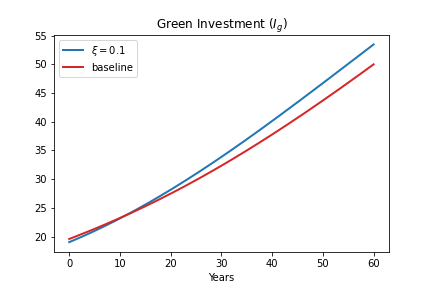}
            \caption[Network2]%
            {{\small Green capital investment}}\label{fig:PreJumpIg}       
        \end{subfigure}
        
        \vspace{0.25cm}
\end{center}        

%\begin{normal}
\caption{
%Figure \ref{fig:ems_rd_inv} shows the 
Simulated outcomes under different uncertainty penalty configurations based on the numerical solutions. Panel (a) shows the pathway for carbon emissions. Panel (b) shows the pathway for R\&D investment as a fraction of total output. Panel (c) shows the level of dirty capital investment. Panel (d) shows the level of green capital investment. The trajectories are simulated under the baseline probabilities abstracting from the intrinsic randomness. The pathways stop after 30 years of simulated outcomes, near the time that the temperature anomaly reaches $1.5^{\circ} C$.
}\label{fig:ems_rd_inv}
%\end{normal}  

\end{figure}

%%%%%%%%%%%%%%%%%%%%%%%%%%%%%%%%%%%%%%%%%%%%%%%%%%%%%%

We focus on the simulated pathways from the pre-damage jump and pre-technology jump state for emissions, R\&D investment, green and dirty capital investment, as well the distribution of climate and damage models, and the probabilities of a damage or technology jump occurring. We examine each of these results across different uncertainty aversion parameters to highlight the implications of increased model uncertainty on the optimal policy choices by our planner. The red lines, or red histogram bars, represent the uncertainty neutral case when $\xi = \infty$, and the blue lines, or blue histogram bars, represent the uncertainty averse case when $\xi = 0.1$. 

%Placeholders, actual figures and discussion of outcomes to come here. We'll want plots of similar outcomes: simulated pathways for emissions, investment, R\&D, distorted probabilities for various model dimensions, and possibly social valuations.

Figure \ref{fig:ems_rd_inv} provides the main economic policy outcomes of interest. Figures \ref{fig:PreJumpEms} and \ref{fig:PreJumpRD} show the pathways for emissions and R\&D investment across the uncertainty averse and uncertainty neutral cases. First, in the top left panel we see that emissions in each case start out similarly, and though they are increasing over time for both cases, we can see that the emissions diverge and are lower for the uncertainty averse case of $\xi = 0.1$. In the top right panel, we see R\&D investment as a percent of total output. Note first that the magnitude of R\&D is fairly substantial in each case, ranging between $2.5\%$ and $3.5\%$ of total output initially. For comparison's sake, \cite{stine2008manhattan} notes that expenditure on R\&D directed towards the Manhattan, Apollo, and the Federal Energy Technology programs reach magnitudes near $0.5\%$, with total R\&D levels reaching over $2\%$. We can also see that when the planner is concerned about model uncertainty, i.e., for $\xi = 0.1$, the fraction of output committed to R\&D investment starts substantially higher, and though the R\&D investment beings to decrease before the uncertaintry neutral case, it always remains higher when there are concerns about model uncertainty. 

 Figures \ref{fig:PreJumpId} and \ref{fig:PreJumpIg} show the levels of dirty and clean investment across the uncertainty aversion cases. Comparing the dirty investment in the bottom left panel and the clean investment in the bottom right panel, we see three key points. First, the level of investment in green capital is substantially higher than in dirty capital. At the beginning of the simulation pathway, the green investment level is about $50\%$ higher, but by the end of the pathway, it is nearly four to five times higher. This highlights the second key observation related to investment, that green investment persistently increases during the simulation pathway, whereas dirty investment begins to taper off and diminish over time. Finally, we see that uncertainty concerns lead to initially higher dirty investment, that decreases more rapidly than in the uncertainty neutral case, whereas there is only a slight amplification of green investment due to difference in uncertainty aversion.

\begin{figure}[!pht]

%\vspace{-1.5cm}
%\vspace{-0.5cm}

%\caption{Numerical Results - Distorted Distributions and Probabilities}  \label{fig:uncertainty}
\begin{center}

        \begin{subfigure}[b]{0.495\textwidth}  
            \centering 
            \includegraphics[width=\textwidth]{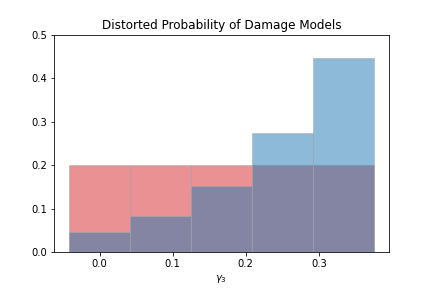}
            \caption[Network2]%
            {{\small Gigatons of carbon emissions}}\label{fig:DmgHist}           
        \end{subfigure}
        \hfill
        \begin{subfigure}[b]{0.495\textwidth}
            \centering
            \includegraphics[width=\textwidth]{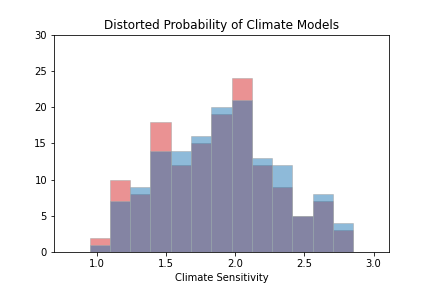}
            \caption[]%
            {{\small R\&D investment-to-output ratio}}\label{fig:ClimateHist}                 
        \end{subfigure}
        
        \vskip\baselineskip
        
        \begin{subfigure}[b]{0.495\textwidth}   
            \centering 
            \includegraphics[width=\textwidth]{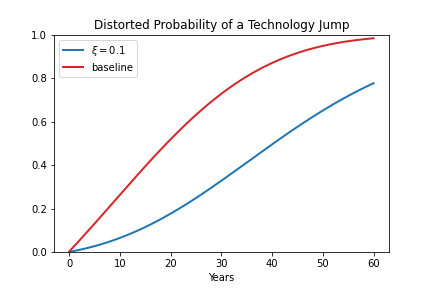}
            \caption[]%
            {{\small Dirty capital investment}}\label{fig:TechJumpProb}    
        \end{subfigure}
                \hfill
        \begin{subfigure}[b]{0.495\textwidth}
            \centering
            \includegraphics[width=\textwidth]{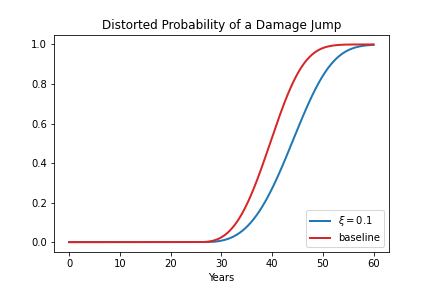}
            \caption[Network2]%
            {{\small Green capital investment}}\label{fig:DmgJumpProb}       
        \end{subfigure}
        
        \vspace{0.25cm}
\end{center}

%\begin{normal}
\caption{
%Figure \ref{fig:uncertainty} shows the 
Distorted model distributions and distorted probability of a jump under different uncertainty penalty configurations based on the numerical solutions. Panel (a) shows the distorted distribution of damage models. Panel (b) shows the distorted distribution of climate models. Panel (c) shows the distorted probability of a technology jump occurring. Panel (d) shows the distorted probability of a damage jump occurring. The trajectories are simulated under the baseline probabilities abstracting from the intrinsic randomness. The histograms are calculated at year 40, and the pathways stop after 40 years of simulated outcomes.
}\label{fig:uncertainty}
%\end{normal}  

\end{figure}

%%%%%%%%%%%%%%%%%%%%%%%%%%%%%%%%%%%%%%%%%%%%%%%%%%%%%%

From Figure \ref{fig:uncertainty} we can see why the planner responds the way they do. The histograms in Figures \ref{fig:DmgHist} and \ref{fig:ClimateHist} show the baseline and distorted probabilities given to damage models (left panel) and climate models (right panel) by the planner. For each type of model, the planner adjusts the probability to give more weight to the right end of the distribution where the implied level of climate damage and climate change are both more severe. The effect is fairly modest for the climate models, and somewhat more pronounced for the damage models. Thus, the planner has an increased incentive to move away from dirty capital and towards green capital when concerns about model uncertainty are present. The pathways in Figures \ref{fig:TechJumpProb} and \ref{fig:DmgJumpProb} show the baseline and distorted probability of a technology jump (left panel) and damage jump (right panel) taking place by the planner. We see only a small adjustment to the jump probability for damages. However, the uncertainty averse planner significantly down-weights the probability of a technology jump occurring when they incorporate concerns about model uncertainty. This adjustment is the larger distributional impact in relative terms, and highlights how the optimal endogenous response by the planner for determining robust policy is to focus their model uncertainty concerns on the technology jump. However, rather than reduce their R\&D investment, the planner emphasizes even further the technological change channel and increases their R\&D investment to try and increase the likelihood of the innovation shock occurring. 

%Our numerical results are therefore able to identify two key results. First is the prominence of technological innovation for policymakers or social planners when considering optimal climate policy and model uncertainty. The second is that the type of technological change that is expected to take place in the future is a critical determinant of how uncertainty about technological change influences the optimal choice of R\&D investment by a social planner.

In summary, it is clear that climate concerns lead to substantial policy action, where the social planner allocates significant resources to R\&D investment and green capital investment, while diverting some of those resources away from dirty investment. However, concerns about uncertainty aversion have relatively modest impacts on investment choices in clean production capital, instead focusing somewhat more on reducing dirty production capital and leaning even more heavily into the policy response of amplifying R\&D investment. The social optimality of this response is driven by the potentially significant payoff from technological innovation in this setting, the fact that emissions are sticky in the sense that they are proportional to output from production using dirty capital and there is no direct mechanism for removing or transforming capital, and the costly nature of accumulating new green capital that is less productive than dirty capital until a technology jump occurs. As a result, the planner places a substantial social valuation on R\&D investment as a policy response tool in order to potentially initiate the technology jump as relying mainly on emissions reductions through additional dirty capital investment reduction or further investment in green capital.

%%%%%%%%%%%%%%%%%%%%%%%%%%%%%%%%%%%%%%%%%%%%%%%%%%%%%%

%Discussion...

%Key points: Big uncertainty concerns over likelihood of technology jump...
%Leads to similar emissions, but significantly lower R\&D. Planner chooses to adjust the dirty investment (down) and green investment (up) instead. Modest adjustment to damage model distribution, little adjustment to the climate models, and almost no adjustment to damage function probability.

%Comparative statics not show:
%Smaller tech. shock means bigger concerns about damage models.
%Bigger tech shock amplifies tech jump concerns, amplifies investment decisions leading to noticeably larger reduction in emissions.
%Post jump? Can we look at this?

\subsection{Validation of Neural Network Solutions}\label{sec:validation}

\begin{figure}[!pht]

%\vspace{-1.5cm}
%\vspace{-0.5cm}

%\caption{Numerical Results - Model Solution Verification}  \label{fig:validation}
\begin{center}

        \begin{subfigure}[b]{\textwidth}  
            \centering 
  \includegraphics[scale=0.95]{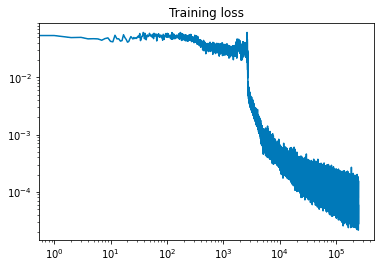} 
            \caption[]%
            {{\small HJB Error for post-technology, post-damage jump state solution}}\label{fig:training_loss}           
        \end{subfigure}

        \vskip\baselineskip
        
        \begin{subfigure}[b]{\textwidth}   
            \centering 
  \includegraphics[width=1.0\textwidth]{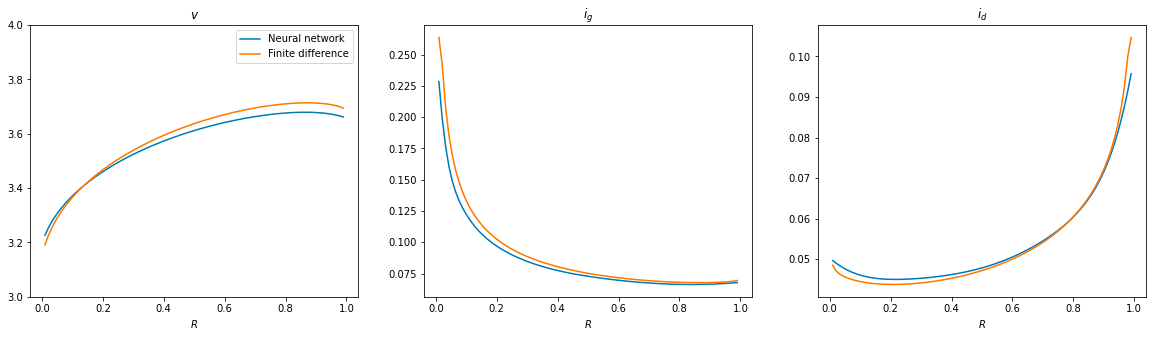} 
            \caption[]%
            {{\small Finite difference and DGM-PIA solution comparison}}\label{fig:FD_compare}    
        \end{subfigure}

        \vspace{0.25cm}
\end{center}        

% \parbox{\textwidth}{\scriptsize Model solution validation. The bottom figure plots the training error over epochs. The bottom figures plot the value function and optimal controls for the FD and NN solution in the post-technology and post-damage jump state.  Still to come: \cite{han2016deep} solution for a given point in the state space.}

%\begin{normal}
\caption{
Model solution validation. The top figure plots the training error over epochs for the post-damage and post-technology jump state. The bottom figures plot the value function and optimal controls for the FD and NN solutions in the post-technology and post-damage jump state.  Still to come is the \cite{han2016deep} solution for a given point in the state space.
}\label{fig:validation}
%\end{normal}  

\end{figure}

 An important issue to address for our numerical solutions is the validation of the accuracy of our results. We address this in three ways, with the results shown in Figure \ref{fig:validation}. First, we examine the training loss for our neural net solutions to determine the magnitude of the HJB equation error for our solution from the DGM-PIA method. Figure \ref{fig:training_loss} shows the training error for the post-technology, post-damage jump state solution across epochs. We are able to reach a training error of approximately $10^{-4}$, suggesting our solution is reasonably accurate.

 In addition, because the post-jump model solutions require only 3 state variables, we can compare and validate these results with solutions derived using the solution derived from more standard finite difference methods. In particular, we use the false-transient, conjugate gradient-based numerical algorithm implemented in \cite{BarnettBrockHansen:2020} and \cite{BarnettBrockHansen:2021} for our finite difference solutions. Figure \ref{fig:FD_compare} shows the neural net and finite difference solutions of the value function and optimal clean and dirty investment for different values of $R$, and values of $\log K$ and $Y$ fixed at the midpoints of our state space range ($\log K = 5.5, Y = 1.5$). We see that the two solutions methods provide consistent outcomes.

 Finally, for the pre-technology, pre-damage jump state, we note that the training errors are similar to the post-damage, post-technology jump state, but we cannot derive finite difference method solutions for this setting. Therefore, we propose the DGM-PIA solution at a given point in the state space to the solution derived from another deep learning method proposed in \cite{han2016deep}, in which the authors directly parameterize the control by neural nets and obtain the optimal parameters by maximizing the approximated utility. This method is known to be very stable, though significantly slower, because the solution is derived point-by-point via training neural nets across repeated simulations of the state variables. We provide these results in future work, though preliminary results show the solutions for the two methods are roughly consistent.

\begin{comment}
\begin{figure}[pht!]
	\centering
% \caption{Distorted Damage and Climate Model Probabilities}\label{fig:PreJumpHistograms}
  \includegraphics[scale=0.95]{figures/Training_Loss.png} 
  \caption{Distorted probability of damage models (left panel) and climate models (right panel). Calculated for the penalty configuration of $\xi = 0.2$. The underlying trajectories are simulated under the baseline probabilities abstracting from the intrinsic randomness.}\label{fig:PreJumpHistograms}
\end{figure}

\vspace{-0.5cm}

\begin{figure}[pht!]
	\centering
  \includegraphics[width=1.0\textwidth]{figures/NN_FD_Comp.png} 
\caption{Distorted probability of damage models (left panel) and climate models (right panel). Calculated for the penalty configuration of $\xi = 0.2$. The underlying trajectories are simulated under the baseline probabilities abstracting from the intrinsic randomness.}\label{fig:PreJumpHistograms}
\end{figure}
\end{comment}

%%%%%%%%%%%%%%%%%%%%%%%%%%%%%%%%%%%%%%%%%%%%%%%%%%%%%%

\newpage
\clearpage

\section{Conclusion}

Model uncertainty along multiple dimensions is a central issue when studying optimal climate policy for a carbon-neutral transition. We integrate dynamic decision theory under uncertainty into a climate-economics-innovation framework with multiple capital stocks to guide uncertainty quantification of optimal R\&D investment, and investment in dirty and clean production capital. We examine different forms of uncertainty, including diffusion and jump process model misspecification, as well as the implications of uncertainty from climate sensitivity, climate damages, and technological innovation. Concerns about model uncertainty feed into the endogenous equilibrium policy responses in our model by ``tilting'' the stochastic discounting implicitly used in constructing the marginal valuations related to the externalities in our framework. This results in first-order implications for the socially optimal outcomes in our model as a result of incorporating aversion to model uncertainty in the planner's decision problem. In particular, uncertainty aversion amplifies the incentive to invest in technological innovation as the major policy mechanism.

Given the richness of our economic and geoscientific model components, we develop and implement a deep learning-based algorithm to solve the Hamilton-Jacobi-Bellman equations that characterize the planner's value function in our setting. The algorithm allows us to derive global solutions for models that have many endogenous and non-stationary state variables, maximize lifetime expected utility over an infinite horizon, are in continuous-time, and have potentially significant non-linearities. Because the computational method can handle such computationally difficult problems without being bogged down by ``the curse of dimensionality'' or other numerical issues, we see significant potential for our algorithm in addressing other economically and geoscientifically rich climate-economic problems. For the same reasons, our computational method also has significant promise for recursive, dynamic general equilibrium models used across economics and finances. We leave exploration of such alternative settings for future work.

%In particular, our toolset is well-suited to handle settings that introduce additional depth relating to optimal policy and model uncertainty concerns such as when there is spatial, risk exposure, and production technology heterogeneities, as well as multi-dimensional climate dynamics with nonlinear thresholds such as those proposed in Chavez, Ghil, and Rombouts (2023).

%%%%%%%%%%%%%%%%%%%%%%%%%%%%%%%%%%%%%%%%%%%%%%%%%%%%%%%%%%%%
%%%%%%%%%%%%%%%%%%%%%%%%%%%%%%%%%%%%%%%%%%%%%%%%%%%%%%%%%%%%

\newpage
\begin{small}
\bibliography{climate,deeplearning}
\end{small} 

\newpage

\begin{appendices}

\setcounter{page}{1}

\section*{Appendix}

This appendix is provided in support of the paper ``A Deep Learning Analysis of Climate Change, Innovation, and Uncertainty'' by Michael Barnett, William Brock, Lars Peter Hansen, Ruimeng Hu, and Joseph Huang. Included here are details on theoretical derivations, alternative model settings, the model parameters, and the numerical methods used in the paper.

\section{Full HJB Equations}

\subsection{No Model Uncertainty Aversion}

\subsubsection{Post Damage and Technology Jumps HJB Equation}

We provide this HJB equation in the main text.

\subsubsection{Intermediate Jump State HJB Equations}
The HJB equation for the pre-damage and post-technology jumps state is given by
\begin{equation*}
\begin{split}
\delta v^{(j)} & =\max_{i_{g},i_{d}}\delta\log([A_{d}-i_{d}](1-R)+[A_{g}'-i_{g}]R ) + \delta\log K\\
 & + \left((1-R) \left[\alpha_{d}+\Gamma_d \log (1+ \theta_d i_{d}) \right]+R \left[\alpha_{g}+\Gamma_g \log (1+ \theta_g i_{g})\right]-\frac{   \sigma_d^2 (1-R)^2 + \sigma_g^2 R^2 }{2}  \right) v^{(j)}_{\log K}\\
 &+ \frac{  \sigma_d^2 (1-R)^2 + \sigma_g^2 R^2}{2} v^{(j)}_{\log K,  \log K}  \\
 & +\left( \left[\alpha_{g}+\Gamma_g \log (1+ \theta_g i_{g})\right] - R \sigma_g^2 - \left[\alpha_{d}+\Gamma_d \log (1+ \theta_d i_{d})\right] + (1-R) \sigma_d^2
  \right) R (1-R) v^{(j)}_{R}\\
 & + \frac{1}{2} R^2 (1-R)^2 ({ \sigma_g^2 + \sigma_d^2} ) v^{(j)}_{RR} \\
  & + \left[-R(1-R)^2\sigma_d^2 + R^2(1-R)\sigma_g^2\right] v^{(j)}_{\log K, R} \\
 & +\beta_{f}\eta A_{d}(1-R)K v_{y}+\frac{|\varsigma|^{2}(\eta A_{d}(1-R)K)^{2}}{2} v^{(j)}_{yy}\\
 & - \left( \{\gamma_{1}+\gamma_{2}Y \}\beta_{f}\eta A_{d}(1-R)K+\frac{1}{2}\gamma_{2} |\varsigma|^{2}(\eta A_{d}(1-R)K)^{2} \right) \\
 & + \mathcal{I}_d(y)  \sum_{m=1}^{M} \pi_d^m ( v^{(m,j)} -  v^{(j)}).
\end{split}
\end{equation*}

The HJB equation for the post-damage and pre-technology jumps state is given by
\begin{equation*}
\begin{split}
\delta v^{(m)} & =\max_{i_{g},i_{d}}\delta\log([A_{d}-i_{d}](1-R)+[A_{g}-i_{g}]R -i_{\kappa}) + \delta\log K\\
 & + \left((1-R) \left[\alpha_{d}+\Gamma_d \log (1+ \theta_d i_{d}) \right]+R \left[\alpha_{g}+\Gamma_g \log (1+ \theta_g i_{g})\right]-\frac{   \sigma_d^2 (1-R)^2 + \sigma_g^2 R^2 }{2}  \right)v^{(m)}_{\log K}\\
 &+ \frac{  \sigma_d^2 (1-R)^2 + \sigma_g^2 R^2}{2} v^{(m)}_{\log K,  \log K}  \\
 & +\left( \left[\alpha_{g}+\Gamma_g \log (1+ \theta_g i_{g})\right] - R \sigma_g^2 - \left[\alpha_{d}+\Gamma_d \log (1+ \theta_d i_{d})\right] + (1-R) \sigma_d^2
  \right) R (1-R) v^{(m)}_{R}\\
 & + \frac{1}{2} R^2 (1-R)^2 ({ \sigma_g^2 + \sigma_d^2} ) v^{(m)}_{RR} \\
  & + \left[-R(1-R)^2\sigma_d^2 + R^2(1-R)\sigma_g^2\right]v^{(m)}_{\log K, R} \\
 & +\beta_{f}\eta A_{d}(1-R)K v^{(m)}_{y}+\frac{|\varsigma|^{2}(\eta A_{d}(1-R)K)^{2}}{2} v^{(m)}_{yy}\\
 & - \left( \{\gamma_{1}+\gamma_{2}\hat{Y}  + \gamma_3^m (\hat{Y} - \underbar{y})  \}\beta_{f}\eta A_{d}(1-R)K+\frac{1}{2}(\gamma_{2} + \gamma_3^m ) |\varsigma|^{2}(\eta A_{d}(1-R)K)^{2} \right) \\
  & +(-\zeta+\psi_{0}i_{\kappa}^{\psi_{1}}\exp(\psi_1 (\log K - \log \kappa))-\frac{1}{2}|\sigma_{\kappa}|^{2})v^{(m)}_{\log\kappa}+\frac{|\sigma_{\kappa}|^{2}}{2}v^{(m)}_{\log\kappa,\log\kappa} \\
 & +\mathcal{I}_g(\kappa)(v^{(m,j)}-v^{(m)}).
\end{split}
\end{equation*}

\subsubsection{Pre Damage and Technology Jumps HJB Equation}

The HJB equation for the pre damage and technology jumps setting is given by

\begin{equation*}
\begin{split}
\delta v & =\max_{i_{g},i_{d}}\delta\log([A_{d}-i_{d}](1-R)+[A_{g}-i_{g}]R -i_{\kappa}) + \delta\log K\\
 & + \left((1-R) \left[\alpha_{d}+\Gamma_d \log (1+ \theta_d i_{d}) \right]+R \left[\alpha_{g}+\Gamma_g \log (1+ \theta_g i_{g})\right]-\frac{   \sigma_d^2 (1-R)^2 + \sigma_g^2 R^2 }{2}  \right)v_{\log K}\\
 &+ \frac{  \sigma_d^2 (1-R)^2 + \sigma_g^2 R^2}{2} v_{\log K,  \log K}  \\
 & +\left( \left[\alpha_{g}+\Gamma_g \log (1+ \theta_g i_{g})\right] - R \sigma_g^2 - \left[\alpha_{d}+\Gamma_d \log (1+ \theta_d i_{d})\right] + (1-R) \sigma_d^2
  \right) R (1-R) v_{R}\\
 & + \frac{1}{2} R^2 (1-R)^2 ({ \sigma_g^2 + \sigma_d^2} ) v_{RR} \\
  & + \left[-R(1-R)^2\sigma_d^2 + R^2(1-R)\sigma_g^2\right]v_{\log K, R} \\
 & +\beta_{f}\eta A_{d}(1-R)K v_{y}+\frac{|\varsigma|^{2}(\eta A_{d}(1-R)K)^{2}}{2} v_{yy}\\
 & - \left( \{\gamma_{1}+\gamma_{2}Y \}\beta_{f}\eta A_{d}(1-R)K+\frac{1}{2}\gamma_{2} |\varsigma|^{2}( \eta A_{d}(1-R)K)^{2} \right) \\
 & +(-\zeta+\psi_{0}i_{\kappa}^{\psi_{1}}\exp(\psi_1 (\log K - \log \kappa))-\frac{1}{2}|\sigma_{\kappa}|^{2})v_{\log\kappa}+\frac{|\sigma_{\kappa}|^{2}}{2}v_{\log\kappa,\log\kappa}\\
 & +\mathcal{I}_g(\kappa)(v^{(j)}-v) + \mathcal{I}_d(y)  \sum_{m=1}^{M} \pi_d^m (v^{(m)} - v).
\end{split}
\end{equation*}

\newpage
\clearpage

\subsection{Full Misspecification Concerns}

\subsubsection{Post Damage and Technology Jumps Setting}

The HJB equation for the post damage and technology jumps setting is given by
\begin{equation*}
\begin{split}
\delta v^{(m,j)} & = \max_{i_{g},i_{d}} \min_{h} \delta\log([A_{d}-i_{d}](1-R)+[A_{g}^j-i_{g}]R) + \delta\log K\\
 & + \left((1-R) \left[\alpha_{d}+\Gamma_d \log (1+ \theta_d i_{d}) \right]+R \left[\alpha_{g}+\Gamma_g \log (1+ \theta_g i_{g})\right]-\frac{   \sigma_d^2 (1-R)^2 + \sigma_g^2 R^2 }{2}  \right)v^{(m,j)}_{\log K}\\
 &+ \frac{  \sigma_d^2 (1-R)^2 + \sigma_g^2 R^2}{2} v^{(m,j)}_{\log K,  \log K}  \\
 & +\left( \left[\alpha_{g}+\Gamma_g \log (1+ \theta_g i_{g})\right] - R \sigma_g^2 - \left[\alpha_{d}+\Gamma_d \log (1+ \theta_d i_{d})\right] + (1-R) \sigma_d^2
  \right) R (1-R) v^{(m,j)}_{R}\\
 & + \frac{1}{2} R^2 (1-R)^2 ({ \sigma_g^2 + \sigma_d^2} ) v^{(m,j)}_{RR} \\
  & + \left( \left( v^{(m,j)}_{\log K} -  R v^{(m,j)}_{R} \right) (1-R) \sigma_d  + \left( v^{(m,j)}_{\log K} + (1-R) v^{(m,j)}_{R} \right)  R \sigma_g \right) \cdot h \\
  & + \left[-R(1-R)^2\sigma_d^2 + R^2(1-R)\sigma_g^2\right]v^{(m,j)}_{\log K, R} \\
 & +(\beta_{f} + \varsigma \cdot h)\eta A_{d}(1-R)K v^{(m,j)}_{y}+\frac{|\varsigma|^{2}(\eta A_{d}(1-R)K)^{2}}{2} v^{(m,j)}_{yy}\\
 & - \left( \{\gamma_{1}+\gamma_{2}\hat{Y} +\gamma_{3}^m(\hat{Y} - \bar{y}) \}(\beta_{f} + \varsigma \cdot h )\eta A_{d}(1-R)K+\frac{1}{2}(\gamma_{2} + \gamma_{3}^m) |\varsigma|^{2}(\eta A_{d}(1-R)K)^{2} \right)  + \xi \frac{|h|^2}{2}.
\end{split}
\end{equation*}

FOC
\begin{eqnarray*}
h   & = &  - \frac{1}{\xi} \left[ \left( v^{(m,j)}_{\log K} -  R v^{(m,j)}_{R} \right) (1-R) \sigma_d + \left( v^{(m,j)}_{\log K} + (1-R) v^{(m,j)}_{R} \right)  R \sigma_g \right] \\
    &   & - \frac{1}{\xi} \varsigma \eta A_{d}(1-R)K \left(v^{(m,j)}_{y} -  \{\gamma_{1}+\gamma_{2}\hat{Y} +\gamma_{3}^m(\hat{Y} - \bar{y}) \}\right). 
\end{eqnarray*}

\subsubsection{Pre Damage and Post Technology Jumps Setting}

The HJB equation for the pre damage and post technology jumps setting is given by
\begin{equation*}
\begin{split}
\delta v^{(j)} & = \max_{i_{g},i_{d}} \min_{f_m, h} \delta\log([A_{d}-i_{d}](1-R)+[A_{g}^j-i_{g}]R -i_{\kappa}) + \delta\log K\\
 & + \left((1-R) \left[\alpha_{d}+\Gamma_d \log (1+ \theta_d i_{d}) \right]+R \left[\alpha_{g}+\Gamma_g \log (1+ \theta_g i_{g})\right]-\frac{   \sigma_d^2 (1-R)^2 + \sigma_g^2 R^2 }{2}  \right)v^{(j)}_{\log K}\\
 &+ \frac{  \sigma_d^2 (1-R)^2 + \sigma_g^2 R^2}{2} v^{(j)}_{\log K,  \log K}  \\
 & +\left( \left[\alpha_{g}+\Gamma_g \log (1+ \theta_g i_{g})\right] - R \sigma_g^2 - \left[\alpha_{d}+\Gamma_d \log (1+ \theta_d i_{d})\right] + (1-R) \sigma_d^2
  \right) R (1-R) v^{(j)}_{R}\\
 & + \frac{1}{2} R^2 (1-R)^2 ({ \sigma_g^2 + \sigma_d^2} ) v^{(j)}_{RR} \\
  & + \left( \left( v^{(j)}_{\log K} -  R v^{(j)}_{R} \right) (1-R) \sigma_d  + \left( v^{(j)}_{\log K} + (1-R) v^{(j)}_{R} \right)  R \sigma_g \right) \cdot h \\
  & + \left[-R(1-R)^2\sigma_d^2 + R^2(1-R)\sigma_g^2\right]v^{(j)}_{\log K, R} \\
 & +(\beta_{f} + \varsigma \cdot h)\eta A_{d}(1-R)K v^{(j)}_{y}+\frac{|\varsigma|^{2}(\eta A_{d}(1-R)K)^{2}}{2} v^{(j)}_{yy}\\
 & - \left( \{\gamma_{1}+\gamma_{2}Y  \}(\beta_{f} + \varsigma \cdot h )\eta A_{d}(1-R)K+\frac{1}{2} \gamma_{2} |\varsigma|^{2}(\eta A_{d}(1-R)K)^{2} \right) + \xi \frac{|h|^2}{2} \\
 & + \mathcal{I}_d(y)  \sum_{m=1}^{M} \pi_d^m f_m (v^{(m,j)} - v^{(j)})  +  \xi \mathcal{I}_d(y) \sum \pi_d^m \left( 1 - f_m + f_m \log f_m \right) .
\end{split}
\end{equation*}

FOC
\begin{eqnarray*}
f_m & = & \exp \left( -\frac{1}{\xi_m} (v^{(m,j)} - v^{(j)}) \right), \\
h   & = &  - \frac{1}{\xi} \left[ \left( v^{(j)}_{\log K} -  R v^{(j)}_{R} \right) (1-R) \sigma_d + \left( v^{(j)}_{\log K} + (1-R) v^{(j)}_{R} \right)  R \sigma_g \right] \\
    &   & - \frac{1}{\xi} \varsigma \eta A_{d}(1-R)K \left(v^{(j)}_{y} -  \{\gamma_{1}+\gamma_{2}Y \}\right) .
\end{eqnarray*}

\subsubsection{Post Damage and Pre Technology Jumps Setting}

The HJB equation for the post damage and pre technology jumps setting is given by
\begin{equation*}
\begin{split}
\delta v^{(m)} & = \max_{i_{g},i_{d},i_{\kappa}} \min_{g_j, h} \delta\log([A_{d}-i_{d}](1-R)+[A_{g}-i_{g}]R -i_{\kappa}) + \delta\log K\\
 & + \left((1-R) \left[\alpha_{d}+\Gamma_d \log (1+ \theta_d i_{d}) \right]+R \left[\alpha_{g}+\Gamma_g \log (1+ \theta_g i_{g})\right]-\frac{   \sigma_d^2 (1-R)^2 + \sigma_g^2 R^2 }{2}  \right)v^{(m)}_{\log K}\\
 &+ \frac{  \sigma_d^2 (1-R)^2 + \sigma_g^2 R^2}{2} v^{(m)}_{\log K,  \log K}  \\
 & +\left( \left[\alpha_{g}+\Gamma_g \log (1+ \theta_g i_{g})\right] - R \sigma_g^2 - \left[\alpha_{d}+\Gamma_d \log (1+ \theta_d i_{d})\right] + (1-R) \sigma_d^2
  \right) R (1-R) v^{(m)}_{R}\\
 & + \frac{1}{2} R^2 (1-R)^2 ({ \sigma_g^2 + \sigma_d^2} ) v^{(m)}_{RR} \\
  & + \left( \left( v^{(m)}_{\log K} -  R v^{(m)}_{R} \right) (1-R) \sigma_d  + \left( v^{(m)}_{\log K} + (1-R) v^{(m)}_{R} \right)  R \sigma_g \right) \cdot h \\
  & + \left[-R(1-R)^2\sigma_d^2 + R^2(1-R)\sigma_g^2\right]v^{(m)}_{\log K, R} \\
 & +(\beta_{f} + \varsigma \cdot h)\eta A_{d}(1-R)K v^{(m)}_{y}+\frac{|\varsigma|^{2}(\eta A_{d}(1-R)K)^{2}}{2} v^{(m)}_{yy}\\
 & - \left( \{\gamma_{1}+\gamma_{2}\hat{Y} + \gamma_{3}^m(\hat{Y}-\bar{y}) \}(\beta_{f} + \varsigma \cdot h )\eta A_{d}(1-R)K+\frac{1}{2}(\gamma_{2} + \gamma_{3}^m)|\varsigma|^{2}(\eta A_{d}(1-R)K)^{2} \right) + \xi \frac{|h|^2}{2} \\
  & +\left(-\zeta+\psi_{0}i_{\kappa}^{\psi_{1}}\exp(\psi_1 (\log K - \log \kappa) )-\frac{1}{2}|\sigma_{\kappa}|^{2} + \sigma_{\kappa} \cdot h \right) v^{(m)}_{\log \kappa}  +\frac{|\sigma_{\kappa}|^{2}}{2}v^{(m)}_{\log\kappa,\log\kappa}\\
 & + \mathcal{I}_g(\kappa) \sum_{j=1}^J \pi_g^{j} g_j \left( v^{(m,j)}-v^{(m)} \right) + \xi \mathcal{I}_g(\kappa) \sum_{j=1}^J \pi_g^{j} \left( 1 - g_j + g_j \log g_j \right).
\end{split}
\end{equation*}

FOC
\begin{eqnarray*}
g_j & = & \exp \left(- \frac{1}{\xi_m} (v^{(m,j)}-v^{(m)}) \right), \\
h   & = &  - \frac{1}{\xi} \left[ \left( v^{(m)}_{\log K} -  R v^{(m)}_{R} \right) (1-R) \sigma_d + \left( v^{(m)}_{\log K} + (1-R) v^{(m)}_{R} \right)  R \sigma_g \right] \\
    &   & - \frac{1}{\xi} \varsigma \eta A_{d}(1-R)K \left(v^{(m)}_{y} -  \{\gamma_{1}+\gamma_{2}\hat{Y} +\gamma_{3}^m(\hat{Y}-\bar{y}) \}\right) \\
    &   & - \frac{1}{\xi} \sigma_{\kappa} v^{(m)}_{\log \kappa}.
\end{eqnarray*}

\subsubsection{Pre Damage and Technology Jumps Setting}

The HJB equation for the pre damage and technology jumps setting is given by
\begin{equation*}
\begin{split}
\delta v & = \max_{i_{g},i_{d},i_{\kappa}} \min_{g_j, f_m, h} \delta\log([A_{d}-i_{d}](1-R)+[A_{g}-i_{g}]R -i_{\kappa}) + \delta\log K\\
 & + \left((1-R) \left[\alpha_{d}+\Gamma_d \log (1+ \theta_d i_{d}) \right]+R \left[\alpha_{g}+\Gamma_g \log (1+ \theta_g i_{g})\right]-\frac{   \sigma_d^2 (1-R)^2 + \sigma_g^2 R^2 }{2}  \right)v_{\log K}\\
 &+ \frac{  \sigma_d^2 (1-R)^2 + \sigma_g^2 R^2}{2} v_{\log K,  \log K}  \\
 & +\left( \left[\alpha_{g}+\Gamma_g \log (1+ \theta_g i_{g})\right] - R \sigma_g^2 - \left[\alpha_{d}+\Gamma_d \log (1+ \theta_d i_{d})\right] + (1-R) \sigma_d^2
  \right) R (1-R) v_{R}\\
 & + \frac{1}{2} R^2 (1-R)^2 ({ \sigma_g^2 + \sigma_d^2} ) v_{RR} \\
  & + \left( \left( v_{\log K} -  R v_{R} \right) (1-R) \sigma_d  + \left( v_{\log K} + (1-R) v_{R} \right)  R \sigma_g \right) \cdot h \\
  & + \left[-R(1-R)^2\sigma_d^2 + R^2(1-R)\sigma_g^2\right]v_{\log K, R} \\
 & +(\beta_{f} + \varsigma \cdot h)\eta A_{d}(1-R)K v_{y}+\frac{|\varsigma|^{2}(\eta A_{d}(1-R)K)^{2}}{2} v_{yy}\\
 & - \left( \{\gamma_{1}+\gamma_{2}Y \}(\beta_{f} + \varsigma \cdot h )\eta A_{d}(1-R)K+\frac{1}{2}\gamma_{2} |\varsigma|^{2}(\eta A_{d}(1-R)K)^{2} \right) \\
  & +\left(-\zeta+\psi_{0}i_{\kappa}^{\psi_{1}}\exp(\psi_1 (\log K - \log \kappa) D )-\frac{1}{2}|\sigma_{\kappa}|^{2} + \sigma_{\kappa} \cdot h \right) v_{\log \kappa}  +\frac{|\sigma_{\kappa}|^{2}}{2}v_{\log\kappa,\log\kappa}\\
 & + \mathcal{I}_g(\kappa) \sum_{j=1}^J \pi_g^{j} g_j \left( v^{(j)}-v \right) + \mathcal{I}_d(y)  \sum_{m=1}^{M} \pi_d^m f_m (v^{(m)} - v) \\
 & + \xi \frac{|h|^2}{2} + \xi \mathcal{I}_g(\kappa) \sum_{j=1}^J \pi_g^{j} \left( 1 - g_j + g_j \log g_j \right)   +  \xi \mathcal{I}_d(y) \sum \pi_d^m \left( 1 - f_m + f_m \log f_m \right) .
\end{split}
\end{equation*}

FOC
\begin{eqnarray*}
g_j & = & \exp \left(- \frac{1}{\xi_m} (v^{(j)}-v) \right), \\
f_m & = & \exp \left( -\frac{1}{\xi_m} (v^{(m)} - v) \right), \\
h   & = &  - \frac{1}{\xi} \left[ \left( v_{\log K} -  R v_{R} \right) (1-R) \sigma_d + \left( v_{\log K} + (1-R) v_{R} \right)  R \sigma_g \right] \\
    &   & - \frac{1}{\xi} \varsigma \eta A_{d}(1-R)K \left(v_{y} -  \{\gamma_{1}+\gamma_{2}Y \}\right) \\
    &   & - \frac{1}{\xi} \sigma_{\kappa} v_{\log \kappa}. 
\end{eqnarray*}

Using the solutions for $g_j$, $f_m$, and $h$ allows for an algebraic simplification of the form:
\begin{equation*}
\begin{split}
\delta v & = \max_{i_{g},i_{d},i_{\kappa}} \min_{g_j, f_m, h} \delta\log([A_{d}-i_{d}](1-R)+[A_{g}-i_{g}]R -i_{\kappa}) + \delta\log K\\
 & + \left((1-R) \left[\alpha_{d}+\Gamma_d \log (1+ \theta_d i_{d}) \right]+R \left[\alpha_{g}+\Gamma_g \log (1+ \theta_g i_{g})\right]-\frac{   \sigma_d^2 (1-R)^2 + \sigma_g^2 R^2 }{2}  \right)v_{\log K}\\
 &+ \frac{  \sigma_d^2 (1-R)^2 + \sigma_g^2 R^2}{2} v_{\log K,  \log K}  \\
 & +\left( \left[\alpha_{g}+\Gamma_g \log (1+ \theta_g i_{g})\right] - R \sigma_g^2 - \left[\alpha_{d}+\Gamma_d \log (1+ \theta_d i_{d})\right] + (1-R) \sigma_d^2
  \right) R (1-R) v_{R}\\
 & + \frac{1}{2} R^2 (1-R)^2 ({ \sigma_g^2 + \sigma_d^2} ) v_{RR} \\
%  & + \left( \left( v_{\log K} -  R v_{R} \right) (1-R) \sigma_d  + \left( v_{\log K} + (1-R) v_{R} \right)  R \sigma_g \right) \cdot h \\
  & + \left[-R(1-R)^2\sigma_d^2 + R^2(1-R)\sigma_g^2\right]v_{\log K, R} \\
 % & +(\beta_{f} + \varsigma \cdot h)\eta A_{d}(1-R)K v_{y}+\frac{|\varsigma|^{2}(\eta A_{d}(1-R)K)^{2}}{2} v_{yy}\\
  & + \beta_{f} \eta A_{d}(1-R)K v_{y}+\frac{|\varsigma|^{2}(\eta A_{d}(1-R)K)^{2}}{2} v_{yy}\\
 & - \left( \{\gamma_{1}+\gamma_{2}Y \}(\beta_{f} + \varsigma \cdot h )\eta A_{d}(1-R)K+\frac{1}{2}\gamma_{2} |\varsigma|^{2}(\eta A_{d}(1-R)K)^{2} \right) \\
  % & +\left(-\zeta+\psi_{0}i_{\kappa}^{\psi_{1}}\exp(\psi_1 (\log K - \log \kappa) D )-\frac{1}{2}|\sigma_{\kappa}|^{2} + \sigma_{\kappa} \cdot h \right) v_{\log \kappa}  +\frac{|\sigma_{\kappa}|^{2}}{2}v_{\log\kappa,\log\kappa}\\
  & +\left(-\zeta+\psi_{0}i_{\kappa}^{\psi_{1}}\exp(\psi_1 (\log K - \log \kappa) D )-\frac{1}{2}|\sigma_{\kappa}|^{2} \right) v_{\log \kappa}  +\frac{|\sigma_{\kappa}|^{2}}{2}v_{\log\kappa,\log\kappa} \\  
& -\frac{1}{2\xi}(v_{\log K}-Rv_{R})^{2}(1-R)^{2}\sigma_{d}^{2}-\frac{1}{2\xi}(v_{\log K}+(1-R)v_{R})^{2}R^{2}\sigma_{g}^{2}	\\  
& -\frac{1}{2\xi}\varsigma^{2}\eta^{2}A_{d}^{2}(1-R)^{2}K^{2}(v_{y}-\{\gamma_{1}+\gamma_{2}Y\})^{2}-\frac{1}{2\xi}\sigma_{\kappa}^{2}v_{\log\kappa}^{2}	  \\
 % & + \xi \frac{|h|^2}{2} + \xi \mathcal{I}_g(\kappa) \sum_{j=1}^J \pi_g^{j} \left( 1 - g_j  \right)   +  \xi \mathcal{I}_d(y) \sum \pi_d^m \left( 1 - f_m  \right). \\
  & + \xi \mathcal{I}_g(\kappa) \sum_{j=1}^J \pi_g^{j} \left( 1 - g_j  \right)   +  \xi \mathcal{I}_d(y) \sum \pi_d^m \left( 1 - f_m  \right).
\end{split}
\end{equation*}

\section{Abatement Model}

Consider allowing for carbon abatement in the following way. First, emissions are given by 
\begin{eqnarray*}
E_{t}	=\lambda_t A_{d}K_{d}(1-\iota_{t}),
\end{eqnarray*}
where $\beta_{t}$ is the emissions intensity of dirty output. Note that $\iota_{t}$ is the choice of abatement and can be solved for as a function of emissions:
\begin{eqnarray*}
\iota_{t}	=1-\frac{E_{t}}{\lambda_{t}A_{d}K_{d}}.
\end{eqnarray*}

The cost of abatement is given as a fraction of the dirty output, given by
\begin{eqnarray*}
J	=[A_{d}K_{d}+A_{g}K_{g}]\phi_{0}(\iota)^{\phi_{1}}.
\end{eqnarray*}

Finally, market clearing gives us the output constraint
\begin{eqnarray*}
C & = & A_{d}K_{d}+A_{g}K_{g}-I_{d}-I_{g}+I_{\kappa}-J \\
 & = & [A_{d}(1-\phi_{0}(1-\frac{E}{\lambda_t A_{d}K_{d}})^{\phi_{1}})-i_{d}]K_{d}+[A_{g}(1-\phi_{0}(1-\frac{E}{\lambda_t A_{d}K_{d}})^{\phi_{1}})-i_{g}]K_{g}-i_{\kappa}K.
\end{eqnarray*}

With that, we can construct the HJB equation for this setting, which is given as
\begin{eqnarray*}
\delta V & = & \max_{i_{g},i_{d},i_{\mathcal{\kappa}},E}\delta\log([A_{d}K_d + A_g K_g] (1-\phi_{0}(1-\frac{E}{\lambda_t A_{d}K_{d}})^{\phi_{1}})-i_{d}K_{d}-i_{g}K_{g}-i_{\kappa}K) -\delta\log N_{t} \\
& & +\{\alpha_{d}+\Gamma_{d}\log(1+\theta_{d}i_{d})\}V_{d}K_{d}+\{\alpha_{g}+\Gamma_{g}\log(1+\theta_{g}i_{g})\}V_{g}K_{g}+\frac{|\sigma_{d}|^{2}K_{d}^{2}}{2}V_{dd}+\frac{|\sigma_{g}|^{2}K_{g}^{2}}{2}V_{gg} \\
& & +\beta_{f}E_{t}V_{Y}+\frac{|\varsigma|^{2}(E_{t})^{2}}{2}V_{YY}+[\{\gamma_{1}+\gamma_{2}Y_{t}\}\beta_{f}E_{t} +\frac{1}{2}\gamma_{2}|\varsigma|^{2}E_{t}^{2}]V_{\log N}+\frac{|\varsigma|^{2}E_{t}^{2}}{2}V_{\log N\log N} \\
& & +(-\zeta+\psi_{0}i_{\kappa}^{\psi_{1}}\exp(\psi_{1}(\log K-\log\kappa))-\frac{1}{2}|\sigma_{\kappa}|^{2})V_{\log\kappa}+\frac{|\sigma_{\kappa}|^{2}}{2}V_{\log\kappa,\log\kappa}. 
\end{eqnarray*}

The FOC for investment and R\&D are given by
\begin{eqnarray*}
0 & = & -\delta([A_{d}(1-\phi_{0}(1-\frac{E}{\lambda_t A_{d}K_{d}})^{\phi_{1}})-i_{d}]K_{d}+[A_{g}(1-\phi_{0}(1-\frac{E}{\lambda_t A_{d}K_{d}})^{\phi_{1}})-i_{g}]K_{g}-i_{\kappa}K)^{-1} \\ 
& & + \Gamma_d \theta_{d} (1+ \theta_d i_{d})^{-1} V_{d} \\
0 & = & -\delta([A_{d}(1-\phi_{0}(1-\frac{E}{\lambda_t A_{d}K_{d}})^{\phi_{1}})-i_{d}]K_{d}+[A_{g}(1-\phi_{0}(1-\frac{E}{\lambda_t A_{d}K_{d}})^{\phi_{1}})-i_{g}]K_{g}-i_{\kappa}K)^{-1} \\ 
& & + \Gamma_g \theta_{g} (1+ \theta_g i_{g})^{-1} V_{g} \\
0 & = & -\delta([A_{d}(1-\phi_{0}(1-\frac{E}{\lambda_t A_{d}K_{d}})^{\phi_{1}})-i_{d}]K_{d}+[A_{g}(1-\phi_{0}(1-\frac{E}{\lambda_t A_{d}K_{d}})^{\phi_{1}})-i_{g}]K_{g}-i_{\kappa}K)^{-1}K \\ 
& & +\psi_{0}\psi_{1}i_{\kappa}^{\psi_{1}-1}\exp(\psi_{1}(\log K-\log\kappa))V_{\log\kappa} \\
0 & = & \delta([A_{d}(1-\phi_{0}(1-\frac{E}{\lambda_t A_{d}K_{d}})^{\phi_{1}})-i_{d}]K_{d}+[A_{g}(1-\phi_{0}(1-\frac{E}{\lambda_t A_{d}K_{d}})^{\phi_{1}})-i_{g}]K_{g}-i_{\kappa}K)^{-1} \\ 
& & \times \frac{\phi_{0}\phi_{1}(A_{d}K_{d}+A_{g}K_{g})}{\lambda_t A_{d}K_{d}}(1-\frac{E}{\lambda_t A_{d}K_{d}})^{\phi_{1}-1} \\ 
& & +\beta_{f}V_{Y}+|\varsigma|^{2}E_{t}V_{YY}+[\{\gamma_{1}+\gamma_{2}Y_{t}\}\beta_{f}+\gamma_{2}|\varsigma|^{2}E_{t}]V_{\log N}+\{\gamma_{1}+\gamma_{2}Y_{t}\}|\varsigma|^{2}E_{t}V_{\log N\log N}
\end{eqnarray*}

Now we redefine the states to get it all correct. We use $\log K=\log(K_{d}+K_{g})$ and $R=\frac{K_{g}}{K_{d}+K_{g}}$. 

Define $dW_{g}$ as the green capital shock and $dW_{d}$ as the dirty capital shock. Having assumed the shocks are independent, then our cross-partial terms should drop out and we end up with the following:
\begin{eqnarray*}
d\log K & = & (1-R)[\alpha_{d}+i_{d}-\frac{\phi_{d}}{2}i_{d}^{2}]dt+R[\alpha_{g}+i_{g}-\frac{\phi_{g}}{2}i_{g}^{2}]dt \\
& & -\frac{1}{2}|\sigma_{d}(1-R)|^{2}-\frac{1}{2}|\sigma_{g}R|^{2}dt
	+(1-R)\sigma_{d}dW_{d}+R\sigma_{g}dW_{g} \\
dR & = & -[\alpha_{d}+i_{d}-\frac{\phi_{d}}{2}i_{d}^{2}]R(1-R)dt+R(1-R)[\alpha_{g}+i_{g}-\frac{\phi_{g}}{2}i_{g}^{2}]dt \\
& & +R(1-R)^{2}|\sigma_{D}|^{2}dt-R^{2}(1-R)|\sigma_{G}|^{2}dt
	-R(1-R)\sigma_{d}dW_{d}+R(1-R)\sigma_{g}dW_{g}
\end{eqnarray*}
The new HJB equation is given by $V(K_{d},K_{g,}Y,\log\mathcal{I}_{g},\log N)=v(K,R,Y_{t},\log\mathcal{I}_{g})-\log N_{t}$:
\begin{eqnarray*}
\delta v & = & \max_{i_{g},i_{d},i_{\mathcal{\kappa}},E}\delta\log([A_{d}(1-R) + A_{g} R ](1-\phi_{0}(1-\frac{E}{\lambda_t A_{d}K(1-R)})^{\phi_{1}}) - i_{d}(1-R) - i_{g} R - i_{\kappa})+\delta\log K \\
& & +[\{\alpha_{d}+\Gamma_d \log (1+ \theta_d i_{d})\}(1-R)+\{\alpha_{g}+\Gamma_g \log (1+ \theta_g i_{g})\}R-(1-R)^{2}\frac{|\sigma_{d}|^{2}}{2}-R^{2}\frac{|\sigma_{g}|^{2}}{2}]v_{\log K} \\
& & +[\{\alpha_{g}+\Gamma_g \log (1+ \theta_g i_{g})\}R(1-R)-\{\alpha_{d}+\Gamma_d \log (1+ \theta_d i_{d})\}R(1-R) \\
& & +|\sigma_{d}|^{2}R(1-R)^{2}-|\sigma_{g}|^{2}R^{2}(1-R)]v_{R} \\
& & +\{\frac{|\sigma_{d}|^{2}}{2}+\frac{|\sigma_{g}|^{2}}{2}\}R^{2}(1-R)^{2}v_{RR}+\{\frac{|\sigma_{g}|^{2}}{2}R^{2}+\frac{|\sigma_{d}|^{2}}{2}(1-R)^{2}\}v_{\log K,\log K} \\
& & -|\sigma_{d}|^{2}v_{\log K,R}R(1-R)^{2}+|\sigma_{g}|^{2}v_{\log K,R}R^{2}(1-R) \\
& & +\beta_{f}E_{t}v_{Y}+\frac{|\varsigma|^{2}(E_{t})^{2}}{2}v_{YY}-[\{\gamma_{1}+\gamma_{2}Y_{t}\}\beta_{f}E_{t}+\frac{1}{2}\gamma_{2}|\varsigma|^{2}E_{t}^{2}] \\
& & +(-\zeta+\psi_{0}i_{\kappa}^{\psi_{1}}\exp(\psi_{1}(\log K-\log\kappa))-\frac{1}{2}|\sigma_{\kappa}|^{2})v_{\log\kappa}+\frac{|\sigma_{\kappa}|^{2}}{2}v_{\log\kappa,\log\kappa} 
\end{eqnarray*}

The FOC for investment and R\&D are given by
\begin{eqnarray*}
0 & = & -\delta([A_{d}(1-R) + A_{g} R ](1-\phi_{0}(1-\frac{E}{\lambda_t A_{d}K(1-R)})^{\phi_{1}}) - i_{d}(1-R) - i_{g} R - i_{\kappa})^{-1} \\
& & +\Gamma_d \theta_{d} (1+ \theta_d i_{d})^{-1}[v_{k}-(1-R)v_{R}] \\
0 & = & -\delta([A_{d}(1-R) + A_{g} R ](1-\phi_{0}(1-\frac{E}{\lambda_t A_{d}K(1-R)})^{\phi_{1}}) - i_{d}(1-R) - i_{g} R - i_{\kappa})^{-1} \\
& & +\Gamma_g \theta_{g} (1+ \theta_g i_{g})^{-1}[v_{k}+(1-R)v_{R}] \\
0 & = & -\delta([A_{d}(1-R) + A_{g} R ](1-\phi_{0}(1-\frac{E}{\lambda_t A_{d}K(1-R)})^{\phi_{1}}) - i_{d}(1-R) - i_{g} R - i_{\kappa})^{-1} \\
& & +\psi_{0}\psi_{1}i_{\mathcal{I}}^{\psi_{1}-1}\exp(\psi_{1}(\log K-\log\kappa))v_{\log\mathcal{I}_{g}} \\
0 & = & \delta([A_{d}(1-R) + A_{g} R ](1-\phi_{0}(1-\frac{E}{\lambda_t A_{d}K(1-R)})^{\phi_{1}}) - i_{d}(1-R) - i_{g} R - i_{\kappa})^{-1} \\
& & \times \frac{\phi_{0}\phi_{1}(A_{d}(1-R)+A_{g}R)}{\lambda_t A_{d}(1-R)}(1-\frac{E}{\lambda_t A_{d}K(1-R)})^{\phi_{1}-1} \\
& & +\beta_{f}(v_{Y}-\{\gamma_{1}+\gamma_{2}Y_{t}\})+|\varsigma|^{2}E_{t}(v_{YY}-\gamma_{2})
\end{eqnarray*}

So far, the model presented here abstracts from jumps and uncertainty. We could introduce jumps as before with different realizations of $\gamma^m_3$ and $A^j_g$ possible, a jump, or jumps, in the value of $\lambda_t$, and jump and diffusion misspecification concerns across the various different channels as outlined in the main text of the paper.

%%%%%%%%%%%%%%%%%%%%%%%%%%%%%%%%%%%%%%%%%%%%%%%%%%%%%%

\section{Social Valuations}

An important component of our analysis can be captured by social valuations, which are shadow prices for the marginal benefit or marginal cost of an additional unit of flows or stocks the various stocks in our model. In particular, the Social Cost of Carbon (SCC), Social Value of R\&D (SVR), Social Value of Green Capital (SVG), and the Social Value of Dirty Capital (SVD), can each be computed from the FOC of the social planner's problem
\begin{eqnarray*}
SCC_t & \propto & 1000  \eta\{-\beta_{f}v_{Y}-|\varsigma|^{2}E_{t}v_{YY}+\{\gamma_{1}+\gamma_{2}Y_{t}\}\beta_{f}+\gamma_{2}|\varsigma|^{2}E_{t}\} \\
SVR_t & \propto & \psi_0 \psi_1 \left(\frac {I_t^j} {\kappa_t} \right)^{\psi_1 - 1}  v_{\log \kappa}  \\
SVG_t & \propto & \Gamma_{g}\theta_{g}(1+\theta_{g}i_{g})^{-1}[v_{\log K}+(1-R)v_{R}] \\
SVD_t & \propto &  \Gamma_{d}\theta_{d}(1+\theta_{d}i_{d})^{-1}[v_{\log K}-Rv_{R}]
\end{eqnarray*}
where the proportionality scaling for each social valuation term is the inverse of the marginal utility of consumption:
\begin{eqnarray*}
(MUC)^{-1} & = & \delta^{-1} ([A_{d}-i_{d}](1-R)+[A_{g}-i_{g}]R-i_{\mathcal{I}}).   
\end{eqnarray*}

Following \cite{BarnettBrockHansen:2020} and \cite{BarnettBrockHansen:2021}, we can decompose the contribution to these social valuations coming from uncertainty of different forms and from different sources. The decomposition requires solving Feynman-Kac equations that represent the expected discounted value of marginal contributions to each stock variable, where the expectation is varied across the different distributions of the potential models under consideration in our setting, e.g., the baseline prior distribution or various forms of the distorted, uncertainty-adjusted distribution of models. We plan to explore these important valuations in future work on this project.

%\begin{figure}[H]
%	\centering
%		\includegraphics[width=0.45\textwidth]{figures/SCCt_1p5.pdf}
 %       \includegraphics[width=0.45\textwidth]{figures/SCCt_1p5.pdf} \\
%		\includegraphics[width=0.45\textwidth]{figures/SCCt_1p5.pdf}
 %       \includegraphics[width=0.45\textwidth]{figures/SCCt_1p5.pdf}        
%\vspace{0.3cm}

%\parbox{\textwidth}{\scriptsize SCC, SVRD, SVG, and SVD. The trajectories are simulated under the baseline probabilities abstracting from the intrinsic randomness. The pathways stop when the temperature anomaly reaches $1.5^{\circ} C$.\label{fig:PreJumpEmissions}}
%\end{figure}

% \begin{comment}
    
%%%%%%%%%%%%%%%%%%%%%%%%%%%%%%%%%%%%%%%%%%%%%%%%%%%%%%

\section{Alternative Model Parameterization}

%%%%%%%%%%%%%%%%%%%%%%%%%%%%%%%%%%%%%%%%%%%%%%%%%%%%%%
    
Economic Framework Parameters (Log Adjustment Costs - BBH RFS):
\begin{table}[H]
	\centering
	\begin{tabular}{ll}
		Parameters & values\\
		\toprule
		$\delta$ & 0.01\\
		$(\alpha, \Gamma, \theta, \sigma)$ & ( -0.035, 0.060, 16.7, \{0.01, 0.016, 0.02\} )\\
		$(\alpha', \Gamma', \theta', \sigma')$ & ( -0.038, 0.0633, 15.7895, \{0.01, 0.016, 0.02\}) \\
		$(\zeta, \psi_0, \psi_1, \sigma_{\kappa})$ & (0, 0.10583, 0.5, 0.0078)\\
        $\varrho$ & 1120 \\
		$(A_d, A_g; \{A_g^j\})$ & $(0.12, 0.10; \{0.15, 0.20, 0.30\})$ \\
        $(K_d, K_g, Y, \kappa)$ & $0.5 \times (85/0.11), 0.5 \times (85/0.11), 1.1, 11.2)$ \\
%		$ \{A_g^j\} $ & $\{0.15, 0.20, 0.30\}\\
		\bottomrule
	\end{tabular}
\end{table}

\begin{itemize}
	\item $\delta$ is consistent with the subjective discount rate used in macro-asset pricing literature.
	\item $(\Gamma_d, \theta_d)$ and $(\Gamma_g, \theta_g)$ are chosen so that the no-climate, one capital version of the model satisfies three conditions as in \cite{BarnettBrockHansen:2020}:
 \begin{eqnarray*}
     1 & = & \Gamma \theta = 1 \\
     0.02 & = & \alpha + \Gamma \log (1+\theta i) = \mathbb{E}[dK/K]\\
     2.5 & = & \frac{A-i}{\delta} v_{\log K} = \frac{1 + \theta i}{\Gamma \theta } = q
 \end{eqnarray*}
	\item $\sigma_d$ and $\sigma_g$ are chosen to match the values used in the World Bank database.
    \item $A_d, A_g, A_g^j$ generate output consistent with World Bank World GDP values, and allow for a meaningful technological change upon realization of the technology jump.
    \item $(\zeta, \psi_0, \psi_1, \varrho)$ values are based on BLS Total  US R\&D stocks values (scaled to World values), \cite{lucking2019have} and \cite{bloom2019toolkit} estimates for the returns to R\&D investment,   
    and consistent with an expected arrival time of potential carbon neutral technology innovation occurring between 30 and 80 years based on current knowledge levels.
    \item $\sigma_{\kappa}$ is chosen to match the volatility of the other capital stocks.
\end{itemize}

% \end{comment}

%%%%%%%%%%%%%%%%%%%%%%%%%%%%%%%%%%%%%%%%%%%%%%%%%%%%%%

%Discussion...

\section{Neural Nets Implementations}

Below we give the pseudo-code for the deep Galerkin method - policy improvement algorithms (DGM-PIA) described in Section~\ref{sec:DGM-PIA}, as well as implementation details, including network architectures and training hyperparameters. 

For the DGM-PIA described in Section~\ref{sec:DGM-PIA}, we use feedforward neural networks with 4 hidden layers of width 32 and \texttt{tanh} activation function (except for the output layer) to approximate both the unknown value functions and the optimal controls. At the output layer, a customized hyperbolic tangent function is used for $i_g, i_d$, and the \texttt{sigmoid} function is used for $i_\kappa$.

To train the neural nets, we run $250000$ epochs with a batch size of $32$. The learning rates $\mathrm{lr}_V = 10e^{-5}$ and $\mathrm{lr}_{\bm \alpha} = 10e^{-5}$ and we use the ADAM optimizer proposed in \cite{kingma2014adam}. 
The training is done on Google Colab with a total runtime of 2.29 hours. 

For the \cite{han2016deep} method solution, we use feedforward neural networks with 4 hidden layers of size 32 and \texttt{tanh} activation function to approximate the optimal controls. To train the neural nets, we run N = 100,000 epochs with a batch size of $2^7$. The infinite time horizon is approximated by $[0, 1000]$, and the time step size is $0.1$. 
%mentioned in Section~\ref{sec:validation}

%The algorithm is implemented on XXX and the total runtime is XXX seconds.

%\jh{Still running so we don't know the run time yet}

\vspace{1.0cm}

\begin{algorithm}
    \caption{The DGM-PIA algorithm for solving the generic HJB \eqref{eq:HJB_generic}}
  \begin{algorithmic}[1]
  \Require a maximum number of epochs $N$, initial neural networks for the value function $V(\bm x; \theta_0)$ and the control $\bm\alpha(\bm x; \varphi_0)$, learning rates for the value function and the control $\mathrm{lr}_V$, $\mathrm{lr}_{\bm\alpha}$
     \For{$n= 0, 1, 2, \dots, N$}
        
        \State Generate random samples $\{\bm x_m\}_{m=1}^M$ from the domain $\Omega$ according to $\nu$
        
        \State Compute the value function loss in \eqref{eq:DGM_HJB} using samples $\{\bm x_m\}_{m=1}^M$:
        \begin{equation*}
        L_V\left(\theta_n; \{\bm x_m\}_{m=1}^M\right) = \frac{1}{M}\sum_{m=1}^M \left[-\delta V(\bm x_m; \theta_n) + \mc{L}^{\bm \alpha(\bm x_m; \varphi_n)} V(\bm x_m; \theta_n) + f(\bm x_m, \bm \alpha(\bm x_m; \varphi_n))\right]^2
        \end{equation*}
        
        \State Take a gradient descent step to update $\theta_{n+1}$:
        \begin{equation*}
        \theta_{n+1} = \theta_n - \mathrm{lr}_V \nabla_\theta L_V(\theta_n; \{\bm x_m\}_{m=1}^M)
        \end{equation*}
        
        \State Compute the control loss in \eqref{eq:DGM_ctrl} using samples $\{\bm x_m\}_{m=1}^M$:
        \begin{equation*}
            L_{\bm \alpha}\left(\varphi_n; \{\bm x_m\}_{m=1}^M\right)  = - \frac{1}{M} \sum_{m=1}^M \left[\mc{L}^{\bm \alpha(\bm x_m; \varphi_n)} V(\bm x_m; \theta_n) + f(\bm x_m, \bm \alpha(\bm x_m; \varphi_n))\right]^2
        \end{equation*}

        \State Take a gradient descent step to update $\varphi_{n+1}$:
        \begin{equation*}
            \varphi_{n+1} = \varphi_n - \mathrm{lr}_{\bm \alpha}\nabla_\varphi L_{\bm \alpha}(\varphi_n; \{\bm x_m\}_{m=1}^M)
        \end{equation*}
     \EndFor
  \end{algorithmic} 
  \label{alg:game}
\end{algorithm}

\end{appendices}

\end{document}